\documentclass[11pt,letterpaper]{article}
\pdfoutput=1

\usepackage{graphicx,array}
\usepackage[dvipsnames]{xcolor}
\definecolor{darkblue}{rgb}{0,0.1,0.5}
\definecolor{darkgreen}{rgb}{0,0.5,0.2}
\definecolor{darkred}{RGB}{153,26,0}
\usepackage{ulem}
\definecolor{seablue}{rgb}{0,0.2,0.6}
\usepackage{latexsym}
\usepackage{amssymb, amsmath}
\usepackage{bm}        
\usepackage[numbers,sort&compress]{natbib}
\usepackage{bm,relsize}
\usepackage{slashed}
\usepackage{tikzfeynman}
\definecolor{viola}{RGB}{134,41,198}
\usepackage{mathrsfs}
\usepackage{hyperref} 
\hypersetup{
    colorlinks=true,       
    linkcolor=darkblue,          
    citecolor=BrickRed,        
    filecolor=magenta,      
    urlcolor=MidnightBlue           
}
\usepackage[all]{hypcap} 
\usepackage{subcaption}

\usepackage[font=small,labelfont=bf,labelsep=period]{caption}

\newcommand{\be}{\begin{equation}}
\newcommand{\ee}{\end{equation}}

 \date{\today}

\begin{document}

\begin{flushright}

\end{flushright}
\vspace{.6cm}
\begin{center}
{\LARGE \bf 
Dark Nucleosynthesis:\\
Cross-sections and Astrophysical Signals
}\\
\bigskip\vspace{1cm}
{
\large Rakhi  Mahbubani$^{a,b}$, Michele Redi$^{c,d}$, Andrea Tesi$^{c,d}$}
\\[7mm]
 {\it \small
$^a$ School of Physics, Astronomy and Mathematics, \\University of
Hertfordshire, Hatfield, Hertfordshire, AL10 9AB, UK\\
$^b$ Theoretical Physics Department, CERN, 1211 Geneva 23, Switzerland\\
$^c$INFN Sezione di Firenze, Via G. Sansone 1, I-50019 Sesto Fiorentino, Italy\\
$^d$Department of Physics and Astronomy, University of Florence, Italy
 }

\end{center}

\bigskip \bigskip \bigskip \bigskip

\centerline{\bf Abstract} 
\begin{quote}
We investigate dark matter bound-state formation and
its implication for indirect-detection experiments.
We focus on the case where dark matter is a baryon of a strongly-coupled dark sector and 
provide generic formulae for the formation of shallow nuclear bound states on emission of photons, and W and Z gauge bosons. 
These processes can occur via electric and magnetic transitions, and
give rise to indirect signals that are testable
in monochromatic and diffuse photon measurements by Fermi and HESS.
We also study the validity of factorizing the bound-state formation cross section into a
short-distance nuclear part multiplied by Sommerfeld-enhancement
factors.  
We find that the short-distance nuclear potential often
violates factorization, modifying in particular the location of the
peaks associated with zero-energy bound states.
Finally we revisit bound-state formation of a (weakly-coupled)
Minimal DM quintuplet including isospin-breaking effects, and find it 
gives rise to indirect-detection signals that are compatible with current bounds.

\end{quote}

\vfill
\noindent\line(1,0){188}
{\scriptsize{ \\ E-mail:\texttt{\href{mailto:rakhi@rakhi@cern.ch}{rakhi@cern.ch}, \href{mailto:michele.redi@fi.infn.it}{michele.redi@fi.infn.it}, \href{andrea.tesi@fi.infn.it}{andrea.tesi@fi.infn.it}}}}
\newpage

\tableofcontents

\setcounter{footnote}{0}


\section{Introduction}

Bound states are a critical ingredient of our universe, and of cosmology in particular. 
It is a natural possibility that Dark Matter (DM) particles might also
bind into more complex structures. Indeed if DM is subject to attractive forces, bound states exist in
the spectrum. The formation of these bound states can have consequences for the DM abundance and
indirect-detection signals, as well as the formation of multi-component DM.

In our universe two types of bound states play an important role: hydrogen atoms and nuclei. 
While the former is weakly coupled and can be studied with ordinary
perturbation theory, study of the latter requires non-perturbative methods to control the strong interactions.
In this work we will study in detail the analogous problem in the dark
sector: the formation of DM bound states in the strongly-coupled regime.

Many works have considered the possibility that DM is charged under a new
abelian gauge symmetry, which allows the formation of hydrogen-like DM
bound
states, see \cite{CyrRacine:2012fz} for a review. In \cite{Mitridate:2017izz} we undertook
a systematic study of hydrogen-like bound states
that arise in any weakly-coupled abelian or non-abelian gauge theory (see
also \cite{Harz:2018csl}). This effect is relevant for WIMP DM candidates
such as Standard Model (SM) electroweak multiplets, as it
can significantly modify the annihilation
cross-section that determines the DM thermal abundance. For example we
found that bound state formation increases the mass at which an
SU(2) quintuplet reproduces the critical abundance from 9.5 TeV \cite{Cirelli:2007xd} to
$M\approx14$ TeV.  It also leads to novel indirect-detection signals
due to the emission of quanta in the formation process.

The production of strongly-coupled DM bound states was first studied quantitively in \cite{Redi:2018muu}. 
In this work residual strong interactions support the formation of
more complex structures.  The
canonical example is a scenario where DM is a baryon of a
confining dark
sector \cite{Antipin:2015xia} (see \cite{Kribs:2016cew} for a review). The dynamics is similar to that of QCD,
so nucleons and nuclei may be likewise stable due to a conserved
(dark) baryon number. Computing the relevant nuclear cross sections
seems unfeasible due to the strongly-coupled nature of the problem.
Fortunately, as was understood and exploited in nuclear physics, this is not the case if the bound states are shallow. As shown by Bethe and Longmire in the '50s
\cite{Bethe:1950jm} and systematically rederived using nuclear
effective theories in the '90s \cite{Kaplan:1998tg,Savage:1998ae}, the
phenomenology of the bound state formation depends only weakly on the details of the underlying potential once the binding
energy is fixed. This observation allows us to reliably compute the
cross section in a controlled effective-theory expansion.
These results can be applied to dark sectors 
with strong interactions such as \cite{Krnjaic:2014xza,mccullough2,mccullough1,Hardy:2014mqa,Hardy:2015boa}

In this work we refine and extend previous studies in several directions. First we address the issue of factorization
of the cross section into long-distance effects (Sommerfeld
enhancement) and a short-distance nuclear cross section.
We show that standard factorization is often violated because the zero-energy bound states associated to Sommerferld peaks 
are modified by the short-distance potential. This effect can be taken
into account in the cross-section computation by including both
long- and short-distance contributions to the potential in
computing the wavefunctions that determine the overlap. 
We present general formulae for bound-state formation
by electric and magnetic emission of electroweak gauge bosons.

Next we consider indirect-detection signals associated to bound-state
formation \cite{Mahbubani:2019pij} (see \cite{Pearce:2013ola,Foot:2014uba,
  Pearce:2015zca,Hardy:2014dea} for weakly-coupled realizations). If
DM is a dark baryon, dark deuterium can be formed through emission of
quanta of energy equal to the binding energy of the state. In the
simplest case where DM is also charged under the SM this is
automatically realized through the emission of electroweak gauge
bosons. This process gives rise to monochromatic photon lines that are
constrained by observation of the Galactic center, as well as a
diffuse photon signal from emission of $W$ and $Z$ gauge bosons and
their subsequent radiation.  For example, when the DM baryon is a
triplet of SU(2)$_L$ the cross section can be within the
reach of current experiments such as FERMI and HESS in the vicinity of
its peaks. We also discuss bound-state formation of Minimal Dark
Matter \cite{Cirelli:2005uq}, and find that the associated signals are
consistent with the current bound on the thermal mass from relic density considerations.

This paper is organized as follows. In Section \ref{sec:factorization}
we present our main  results and illustrate them in the context
of a simple U(1) toy model.
In Section \ref{sec:xsec} we provide general formulae for the
formation of two-particle bound states by emission of a gauge boson. 
In Section \ref{sec:darkSU2} we apply our formalism to the simplest dark baryon model: a complex SU(2)$_L$ triplet. We first compute the
rate of cosmological production of `dark deuterium', improving previous estimates by including isospin-breaking effects, 
and then derive astrophysical signals due to bound-state formation.
In Section \ref{sec:mdm} we apply the same tools to Minimal DM: an elementary quintuplet of SU(2)$_L$. We summarise
the result in Section \ref{sec:conclusions}. A series of technical appendices follow.

\section{Formation of shallow bound states}
\label{sec:factorization}

\begin{figure}
\centering
\begin{tikzpicture}[line width=1.5 pt, scale=1.7]
	\node at (-1.7,0.35) {DM};
	\node at (-1.7,-0.35) {DM};
	\draw[] (-1.5,0.35)--(0,0.35);
	\draw[] (-1.5,-0.35)--(0,-0.35);
	\draw[line width=3.4pt, color=black] (0.7,-0.3)--(0,0);
	\draw[vector,color= blue] (0,0)--(0.7,.3);
	\draw[vector,color= blue] (-1.3,-0.35)--(-1.3,0.35);
	\draw[vector,color= blue] (-1.15,-0.35)--(-1.15,0.35);
	\draw[vector,color= blue] (-1.0,-0.35)--(-1.0,0.35);
	\draw[dotted,color= blue] (-0.85,0)--(-0.65,0);
	\draw[vector,color= blue] (-0.56,-0.35)--(-0.56,0.35);
	\draw[fill, color=darkred] (0,0) circle (.4cm);
	\node at (1.55,-0.33) {DM bound state};
	\node at (1.42,.33) {gauge boson};
\end{tikzpicture}
\vspace{-0.2cm}
\caption{
\label{fig:triplet-indirect} DM bound-state formation. When DM is
charged under a long-range force the initial state wavefunction is
strongly distorted before the short distance nuclear process, depicted as red blob, takes place.}
\end{figure}
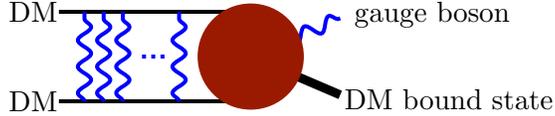

We are interested in computing the formation of bound states, which we
will refer to generically as `dark deuterium', via emission of SM gauge bosons. 
This is always kinematically allowed for photons, while for SU(2)
gauge bosons the binding energy must be larger than their mass in
order for the process to take place at rest.

For the purposes of this work we take DM to be a spin-1/2 dark baryon
triplet, $V$, of SU(2)$_L$. This could arise, for example, from
an $SU(3)_D$ gauge theory with fermions in the fundamental
representation of dark color, which are also triplets of SU(2)$_L$
\cite{Antipin:2015xia}.  The lightest nucleon is the neutral component
of the baryon triplet, which has a $165$-MeV mass splitting from the
charged component due to EW-loop effects, and it is stable due to
dark-baryon-number conservation. Since DM has no electric charge,
bound-state formation by emission of photons cannot take place at tree
level, and requires the inclusion of long-distance effects. If DM is
an electroweak multiplet, Sommerfeld enhancement (SE) 
due to multiple exchanges of electroweak gauge bosons can turn a pair
of $V^0 V^0$ into $V^+ V^-$.  The latter can then emit photons on
bound-state formation. Alternatively the strong interaction can also allow for the same transition. 

\bigskip
The problem of computing bound-state formation in the presence of
long-distance effects such as SM gauge interactions is complicated by
the presence of multiple scales. We wish here to clarify the issue of
factorization of the total cross section, which will often not hold for bound state formation\footnote{A different aspect of lack
  of factorization was studied in \cite{Blum:2016nrz}.}. 

We start by listing the relevant scales of the problem:
\begin{itemize}
\item{$r_{\rm dB}=1/(M v_{\rm rel})$:} the de Broglie wave-length is typically the largest length scale in the problem where the particles can be treated as free.
\item{$r_{\Delta}=1/\sqrt{2M \Delta}$:} the length scale where the splitting between charged and neutral components ($\Delta=$165 MeV) becomes
important. For energies below $2 \Delta$ the $V^+ V^-$ pair does not
exist as an asymptotic state, and its  wavefunction goes to zero at
distances larger than $r_{\Delta}$.  Below this distance scale the
system is, to a good approximation, isospin-symmetric, and the
wavefunction of the charged pair is related to that of the neutral
pair by a Clebsch-Gordan coefficient. 
\item{$r_W=1/M_W$:} $SU(2)$ interactions become relevant and approximately symmetric at distances smaller than  $r_W$. This the scale where the SE starts to build up.
\item{$r_B=(\alpha_2 M)^{-1}$} The `Bohr radius' is the distance at
  which the SE approaches the asymptotic value commonly used in the literature.
\item{$a_{i,f}$:} the scattering lengths of the initial and final
  states of the strong nuclear interaction. This corresponds to the size of the shallow bound states in the initial and final channel.
\item{$r_\pi=1/M_\pi$:} the range of the strong interactions. For shallow bound states such as nuclei,  $a_{i,f}> r_\pi$ and detailed knowledge 
of the microscopic physics  becomes unimportant. Indeed, to leading order the cross section only depends on the scattering length, as first pointed out by Bethe \cite{Bethe:1950jm}. 
\end{itemize}

To first approximation the DM annihilation cross section is assumed to
factorize into a short-distance part and a long-distance part as follows:
\begin{equation}
\sigma_{\rm ann} = {\rm SE} \times \sigma_{\rm short}
\end{equation}
where SE is the Sommerfeld enhancement encoding the distortion of the
initial wavefunction due to exchange of light mediators, while $\sigma_{\rm short}$ is computed from microscopic physics.
This factorization is justified in the case of heavy DM, because the annihilation effectively takes place at a distance $r_{\rm ann}\sim 1/M$, the wavelength of the final states, 
which is much shorter that $r_{\rm dB}$ \cite{Hisano:2004ds, Hisano:2006nn}.
For bound-state formation there are two (related) effects that may
invalidate the factorization. First, for some regions of the
strong-sector parameter space, the bound state
cannot be treated as point-like. Second, for massive mediators the
 cross section has peaks that are associated to zero-energy bound states supported by the long-distance potential. 
 These bound states at threshold are very fragile and can be modified by the short-distance potential. 
Although the latter would also spoil factorization in the case of DM
annihilation, the effect is unavoidable in bound-state formation if
the nucleons have SM charges. We describe each of these effects in greater detail below.

\paragraph{Finite size effects:}
The failure to factorize due to finite size effects is already evident
in the production of hydrogen. 
For the ground state, in the electric dipole approximation and at low velocities one finds (see \cite{Asadi:2016ybp})
\begin{equation}
\begin{split}
(\sigma v_{\rm rel})_\text{hydrogen}= 2\pi \left(\frac {\alpha}{v_{\rm rel}}\right)^3 \times e^{-4} \times \frac {2^6}3 \frac {\pi \alpha^2}{M_e^2}  \frac {M_e}{E_B}\, v_{\rm rel}^2  \,. 
\end{split}
\label{hydrogen}
\end{equation}
This formula is written as a p-wave SE times a short distance
cross section that depends only on the final state. The factor
$e^{-4}\approx 0.018$, arising from the non-trivial overlap of the
wavefunctions of initial and final states, encodes the deviation from
the factorized limit. More generally, if the initial state
feels an attraction of strength $\lambda_i \alpha$ while the bound state is associated to a coupling $\lambda_f \alpha$ the deviation from factorization is given by $e^{-4\lambda_i/\lambda_f}$, \
so that in the limit  $\lambda_f  \gg \lambda_i$ (small Bohr radius) the factorization is recovered.

Similarly, for the formation of deuterium at low energies one finds \cite{Bethe:1950jm}
\begin{equation}
(\sigma v)_{pn \to D +\gamma}=\kappa_1^2 \frac  {8\pi \alpha}{M^5} \gamma_f^3 (1-a_i \gamma_f)^2\,,
\end{equation}
where $a_i\approx-23$ fm is the scattering length of the initial $^1S_0$  channel  and $\gamma_f = \sqrt{E_B M}\approx(5 \rm fm)^{-1}$  is the inverse scattering 
length of the $^3S_1$ final state channel. In the formula above the first term in the parenthesis corresponds to the tree-level process while
the second term encodes long-distance nuclear effects associated to the large scattering length of the initial state,  see \cite{Kaplan:2005es}.

It is clear that factorization should be a good approximation if the size of the bound state is smaller than the range of the long distance interactions.
This corresponds to,
\begin{equation}
r_W \ge r_B \ge a_{i,f}
\end{equation}
When this condition is violated the effect of long-distance physics cannot be captured simply by the value of the wavefunction 
at the origin and a full quantum-mechanical computation is required.

\paragraph{Location of the Sommerfeld peaks:}
The other effect that violates factorization relies on the
presence of a short-range attractive potential in addition to a Yukawa-type potential for the initial state. As we will see this effect is very generic.
It has long been known that zero-energy bound states supported by a
Yukawa potential in the initial state give rise to peaks in the cross section at low velocities.
The positions of these peaks correspond to a set of critical masses $M_*$ determined by the equation
\be
E_W(M_*)=0\,,
\ee
where $E_W$ is the binding energy of the bound state supported by the
Yukawa potential (induced for example by electroweak interactions). 
This tuned condition is potentially sensitive to deformations of the
potential due to other effects. Indeed as we will see, any short-range
potential generically contributes to the binding energy of shallow bound states such that the zero-energy bound state appears 
at a different value of the mass. This violates the naive factorization of long distance and short distance effects.

One way to take this into account would be to compute the
SE using the full potential, including both long- and short-distance
contributions.  While this computation reproduces the positions of the
peaks, it does not take into account the fact that the effect of the short-distance
potential is already partially included in the hard scattering cross section.   
It is possible to improve this estimate by evaluating the SE not at
the origin, but at a distance of the order of the scattering length.
Clearly the safest way to proceed is to compute the cross section
directly using explicit wavefunctions and evaluating the overlap between initial and final state.
Importantly, while this procedure requires the choice of an explicit potential, in the
limit of shallow bound states the result becomes independent of the details of the short-range potential.

\subsection{$\mathrm{U(1)}$ Toy model}
In order to illustrate these effects we first consider a simple U(1)
model. In this scenario DM is composite $i.e$ subject to a
short-range nuclear potential, and
coupled to a vector boson of mass $M_V$.
The static potential between two DM particles with opposite charges can be approximated as
\begin{equation}
V(r)= -\alpha_D \frac {e^{-M_V r}}r + V_N \theta(r_0 -r)\,,
\label{eq:1dpot}
\end{equation}
where $\alpha_D$ is the strength of the dark U(1), and $r_0< 1/M_V$ is the range of the nuclear interaction.
As discussed in \cite{Mahbubani:2019pij}, strongly-coupled models with
a dark photon interaction can be effectively parametrized with this type of potentials. 
When the nuclear potential is switched off ($V_N=0$) the Yukawa potential alone supports a zero-energy s-wave bound state for
\begin{equation}
M_*=\{1.65\,,6.4\,,14.5, \dots\} \frac{M_V} {\alpha_D}\,.
\end{equation}
Let us now turn on an attractive short-distance nuclear potential ($V_N<0$). As we increase $|V_N|$ the zero energy bound state becomes
deeper until
\begin{equation}
V_N\approx -\frac {\pi^2} {4 M r_0^2}\,.
\end{equation}
For such a value the nuclear potential is not a small perturbation on
the original Yukawa potential, but rather the converse.
Indeed when this condition is met the nuclear potential alone supports one
bound state. The long distance potential gives a shift to the binding energy of the nuclear bound state
that can be estimated as
\begin{equation}
\Delta E_B \sim \alpha_D \frac{e^{-a M_W}}a\,,
\end{equation}
where we used the fact that the size of the nuclear wavefunction is
comparable to the scattering length, which in turn is determined 
by $a^{-1}\approx \sqrt{M E_B}$. This shows that corrections to the
nuclear binding energies are small, possibly exponentially suppressed.

We can then ask about the fate of the zero-energy bound state supported by the long distance potential, and perturbed by the nuclear potential. 
Naively one might think that the attractive short-distance potential should make the zero-energy bound state more bound. 
However one must keep in mind that this bound state would be the first excited level of the full potential. On general grounds the wavefunction of the
ground state has zero nodes while that of the first excited level has one. This implies an extra positive contribution to the energy from the kinetic term.
Indeed in our example the bound state is less bound, shifting the peak
of the SE to larger masses.

\begin{figure}[t]
\centering
\includegraphics[width=0.49\textwidth]{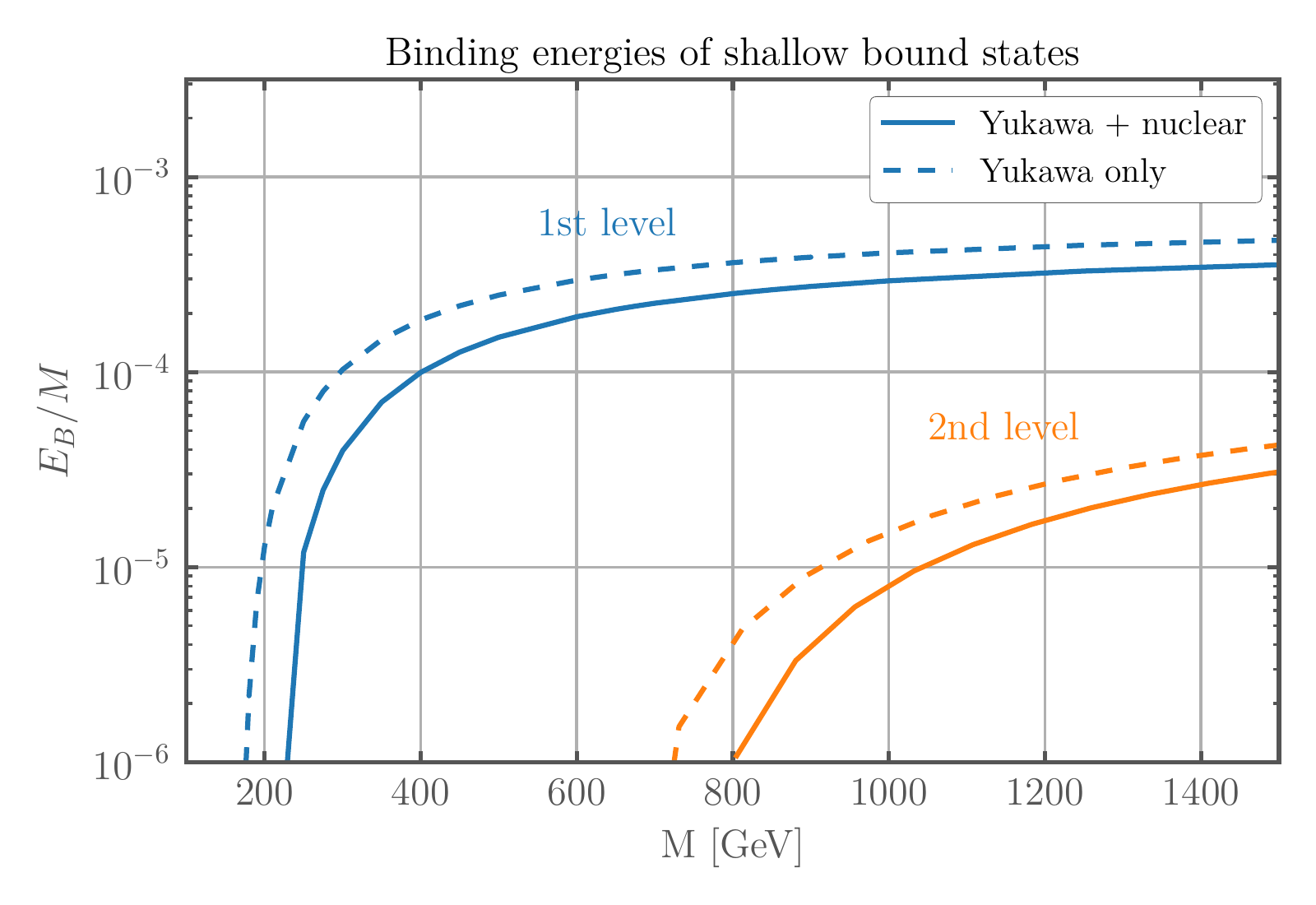}~
\includegraphics[width=0.49\textwidth]{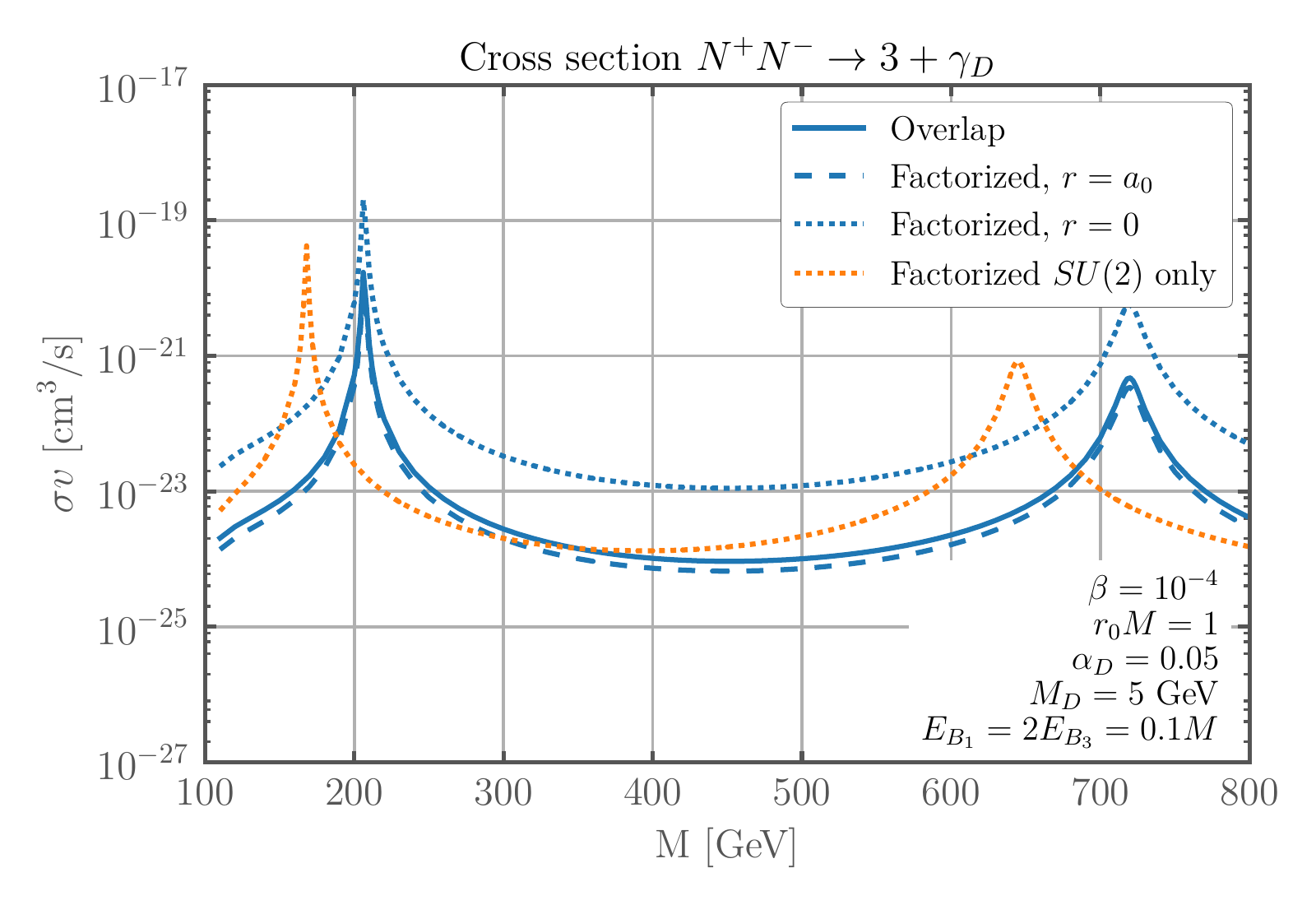}
\caption{\label{fig:binding} Left panel: Binding energies of first two shallow
  bound states in a composite DM model with long-range U(1)
  interactions (solid lines).  The DM potential consists of short-range (nuclear) and
  long-range (Yukawa) components, see \eqref{eq:1dpot}.
  The binding energies with the nuclear
  potential turned off are shown for comparison (dashed lines).
  Right panel: Bound-state formation cross section (solid line) for composite U(1)
 model, showing shift of zero-energy bound states due to the
 short-range nuclear potential.  The factorized cross section is shown
 for comparison, with the SE computed for (i) a long-range Yukawa
 potential only (orange), (ii) Yukawa + short-range nuclear potential,
 evaluated at the origin (blue dotted) and (iii) Yukawa + short-range
 nuclear potential evaluated at $r=a_0$, the size of the initial bound
 state (blue dashed).}
\end{figure}

In Fig.~\ref{fig:binding} we plot the binding energies of the first
two bound states of the full potential (\ref{eq:1dpot}), as well as those of the pure Yukawa potential.
We see that in the presence of a nuclear potential the zero energy
bound state is shifted to larger masses. This implies that the
Sommerfeld peaks of the cross section will appear at different values of the mass. 

For p-wave zero modes the situation is different, the condition for zero-energy p-wave bound states in a spherical well
is $V_N M= \pi^2/r_0^2$; hence for a shallow s-wave nuclear bound
state there is no p-wave nuclear bound state. A zero mode p-wave
state arises solely due to the Yukawa interaction, and it has the
lowest energy level in this sector.  Equivalently, since the
wavefunction vanishes at the origin the effect of the
short-distance potential is small, and the location of the peaks will
not change much. This intuition is confirmed by numerical computations

\begin{figure}[t]
\centering
\includegraphics[width=.6\textwidth]{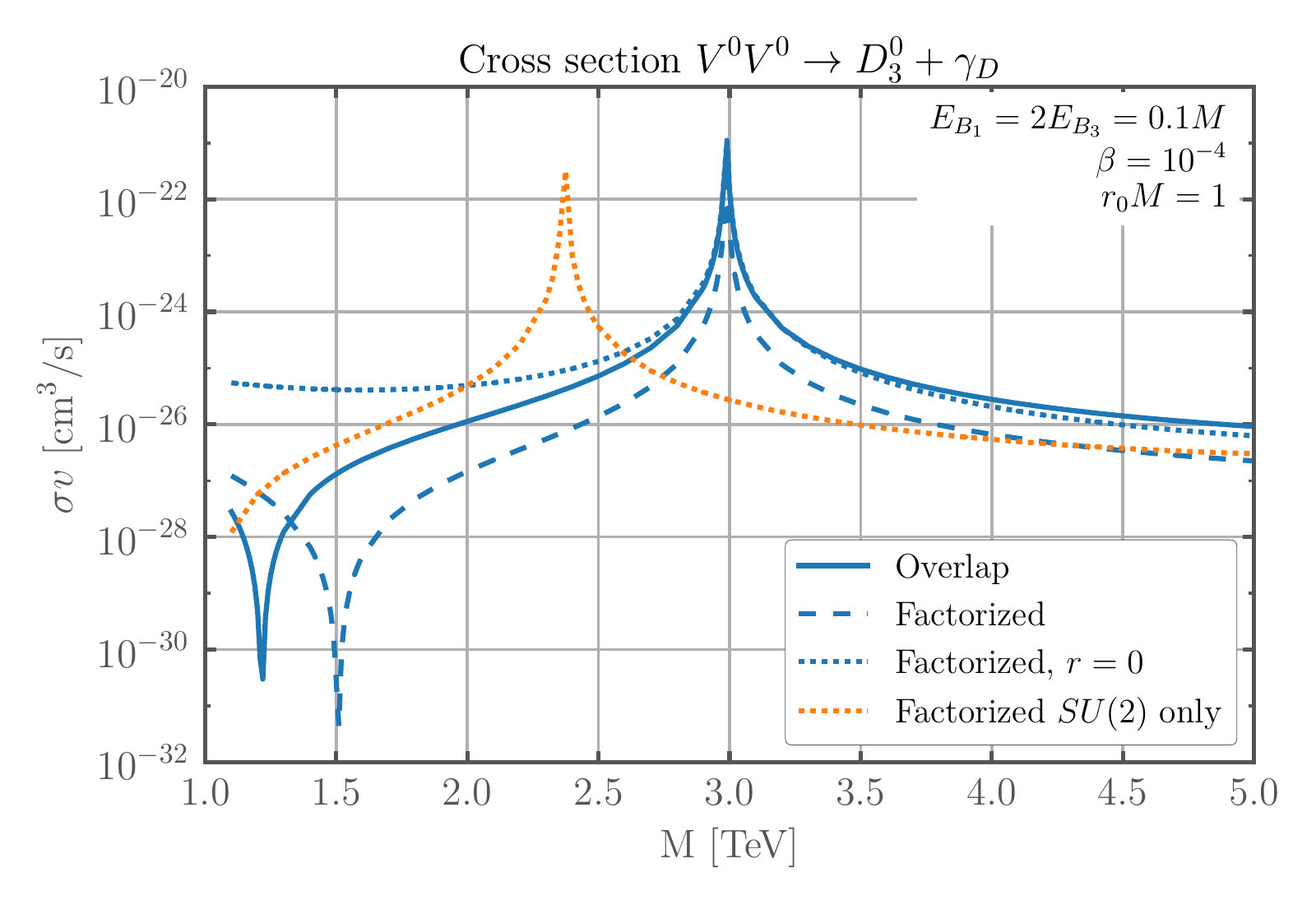}
\caption{\label{fig:BSF1d} Bound state production cross section for
  dark SU(2)$_L$ triplet model computed with
  overlap integrals (blue solid), and factorized cross section with
  SE computed using (i) only the
  long-range component of the potential (orange) (ii) the full
  potential, with SE evaluated
  at the origin (blue dotted) and (iii) the full potential, with SE
  evaluated at $r=a_0$ (dashed blue).}
\end{figure}

Let us now discuss the production of nuclear bound states. For simplicity we focus on magnetic interactions 
which allow the formation of an s-wave bound state from an s-wave initial state
by emission of a dark photon.
As reviewed in Appendix \ref{app:nucl} a shallow s-wave bound state  gives rise to a SE factor,
\begin{equation}
{\rm SE}_0\approx\frac {a_0^2 M V_N}{1+ a_0^2 p^2}\,,
\end{equation}
where $a_0^{-1}\approx \sqrt{M E_B}$ is the s-wave scattering
length, and $p=\sqrt{ME}$ is the momentum of the incoming state (for a
generic potential $V_N$ would be replaced by the typical depth of
the potential). 
These peaks are visible in the total cross section for bound-state formation in
the U(1) toy model.  We compute the cross section using the full quantum-mechanical
overlap, and using the factorized cross section.  Our results are
shown in Fig.~\ref{fig:BSF1d}.
The true cross section is given by the solid blue curve. The factorized
result has a similar shape, with a normalization and shift that
depends on how the Sommerfeld factor was computed.  Including only the
long-range Yukawa potential in the SE computation results in a shift
to lower masses (orange dotted curve) as a consequence of shifted
zero-energy bound states.  The correct location of peaks is captured
by computing the SE with the full potential, containing both
short-range (nuclear) and long-range (Yukawa) components 
\eqref{eq:1dpot}. A very good approximation to the true cross section
is obtained by evaluating the SE at $r=a_0$ (blue dashed) rather than
at the origin (blue dotted).  Note that we only show explicitly here
  the breakdown of factorization due to a shift in the zero-energy
  bound states. In Fig.~\ref{fig:BSF1d} we show the equivalent effect for a dark
  SU(2)$_L$ triplet model, which we consider in greater detail in
  Section \ref{sec:darkSU2} below. In this case the presence of several channels also leads to 
  a dip where the cross section is strongly suppressed.


\section{Cross sections for bound-state formation}
\label{sec:xsec}

In this section we provide generic expressions for bound-state formation, including explicit formulae for
shallow bound states. This extends the work in \cite{Mitridate:2017izz}, where the formation of hydrogen-like
bound states through  electric dipole transitions was studied for general weakly-coupled non-abelian gauge
theories.

As we discussed in the previous section the standard approach to computing the
 relevant cross sections in the case of bound state annihilation,
  {\it i.e.} convolution of the analytic short distance cross section with the
  SE due to long-range forces only, does not work generically.  This is due to
  the shift of the spectrum of zero-energy bound states, as well as
  finite-size effects in some region of parameters. We will thus take a
  numerical approach where we compute the wavefunction
  overlap using the full potential.  While this requires a choice of
  nuclear potential, the results are only weakly dependent on this
  choice in the regime of shallow bound states that will be of
  interest to us here.
  For simplicity we will parametrize the nuclear potential with a spherical well.



A note on calculability is in order. The possibility to compute an intrinsically strongly-coupled process such as nuclei formation
relies on the smallness of nuclear binding energies, allowing us to
describe bound state formation within a local effective theory where the 
pions have been integrated out. The expansion parameter is $p/M_\pi$
where $p=\sqrt{M E}$.  For a bound state this implies
\begin{equation}
\frac {\sqrt{M E_B}}{M_\pi}< 1 \longrightarrow E_B < \frac{M_\pi^2}{M}\,.
\end{equation}
In the SM the expansion parameter is approximately 0.3, hence higher order corrections are of the order 10\%, as seen experimentally.
It remains an open question whether the mild tuning of the binding energy is accidental or a robust feature of nuclear
interactions. The latter possibility is supported by lattice studies \cite{Beane:2012vq}.

\subsection{Quantum mechanical computation}


To compute the cross section in quantum mechanics we follow 
\cite{Mitridate:2017izz}, to which we refer the reader for further details.
 The general process we wish to consider is bound state
  formation through emission of a spin-1 field $\mathcal{V}^a$ with
  mass $M_a$ and momentum $k$:
\begin{equation}
{\rm DM}_i(p_1)+ {\rm DM}_j(p_2)\Longrightarrow B_{i'j'}+ \mathcal{V}^a(k)\,.
\end{equation}
Here the indices $i,j,i',j'$ run over the various components of DM
in the initial and final states,
$B_{i'j'}$ is the bound state, and $a$ runs over the generators of the
adjoint. In the center-of-mass frame the energy of the emitted quantum is
\begin{equation}
\omega=\sqrt{k^2+ M_a^2}\approx E_B + M \beta^2\,,~~~~~~~~~~~~ \beta=\frac{v_{\rm rel}}2\,.
\end{equation}
As usual we can decompose the initial- and final-state wave function in states of fixed orbital angular momentum
\begin{equation}
\psi(r,\theta,\varphi) = \sum_{\ell,m}R_\ell (r)Y_\ell^m(\theta,\varphi)=\sum_{\ell,m}
\frac{u_\ell(r)}{r} Y_\ell^m(\theta,\varphi)
\end{equation}
where $Y_\ell^m$ are spherical harmonics. The radial wave function $u_\ell(r)$  satisfies
\begin{equation}
-\frac {u_\ell''}{M} + \left[ V(r)+ \frac {\ell(\ell+1)}{M r^2}\right] u_\ell = E u_\ell
\end{equation}
where $E=M \beta^2$.  In general this is a matrix-valued equation in the two-particle Hilbert space $|i\rangle \otimes |j\rangle$
and must be solved numerically. Details of our method, including the
boundary conditions and normalizations are given in Appendix \ref{appendixB}.

We will include in the potential a
Yukawa-like long-range component induced by SM interactions (c.f. the
ubiquitous wino in the MSSM) 
as well as a short-range nuclear potential, as in
Eq.~\eqref{eq:1dpot}. For the latter we will use a spherical well with
parameters that reproduce the binding energies we select for the nuclear bound states. 


\paragraph{Magnetic transitions.~\\}
For strongly-coupled bound states, such as the proton and neutron in
the SM, the largest cross section at low velocity is usually due to
magnetic dipole transitions. This observation also holds true in the context of dark
sector bound states, and it can have important cosmological consequences \cite{Redi:2018muu}. The magnetic emission is described in general by the effective Hamiltonian in the two-particle space,
\begin{equation} 
- \frac {\kappa g}{M} \left(T^a_{i'i} \delta_{jj'} \vec{\sigma}\cdot\vec {\mathcal{B}^a}(x_1)+\bar{T}^a_{j'j} \delta_{ii'} \vec{\sigma}\cdot\vec {\mathcal{B}^a}(x_2)\right)
\end{equation}
where the magnetic field
  $\mathcal{B}^{k a}= \epsilon^{klm} F_{lm}^a$ and $\sigma_k$ denotes
  the Pauli matrices. The matrices $T$ and
$\overline{T}$ are generators in the representations of the initial DM states
1 and 2 respectively, of mass $M$. $g$ is the gauge coupling, and the coefficient $\kappa$ is O(1) for strongly-coupled bound states.

For transitions between s-wave states by emission of a gauge boson
$\mathcal{V}^a$ with mass $M_a$ one finds,
\begin{equation}
\begin{aligned}\label{eq:magQM}
(\sigma_{{\rm bsf}} v_{\rm rel})_a^{\rm mag}&=\frac {2^5} {g_N^2} \kappa^2 \alpha \frac{k^3}{M^2} \Bigg|\int r^2drR_{s,ij}\Bigg(\frac{1}{2}\left(T_{i'i}^a\delta_{jj'}-\overline{T}_{j'j}^{a}\delta_{ii'}\right) \Bigg)R^*_{B,j'i'}\Bigg|^2
\end{aligned}
\end{equation}
where $g_N=2(4) d_R$ is the number of degrees of freedom for Majorana
(Dirac) DM and $R_s$ and $R_B$ are the radial wavefunctions for the
initial s-wave state and final bound state, respectively.

If the DM masses are degenerate the theory has an approximate
$SU(N_F)$ flavor symmetry, which is broken by electroweak interactions.
Neglecting this small symmetry breaking effect we can decompose the
wavefunctions in irreducible representations of the global symmetry, $R_{ij}^M(r)= R^M(r) (CG)^M_{ij}$ where $(CG)^M_{ij}$ are the matrices for the change of
basis from $|i\rangle \otimes |j\rangle$ to the irrep labelled by the
index $M$. For electroweak interactions the suggestively named
$(CG)^M_{ij}$ can be chosen to be the Clebsch-Gordan coefficients and
the radial wavefunctions chosen to be real. This choice makes the
Schroedinger equation 1-dimensional, and the magnetic cross section in the isospin basis becomes
\begin{equation}\label{eq:magQMBasis}
(\sigma v_{\rm rel})_{aMM'}^{\rm mag}= \frac {2^5} {g_N^2} \kappa^2 \alpha \frac{ k^3}{M^2} \Bigg|C_J^{a MM'} \times \int r^2drR_s^M R_B^{M'} \Bigg|^2 \, 
\end{equation}
with the group theory factor,
\begin{equation}
C^{aMM'}_J= \frac 1 2 {\rm Tr}[{\rm (CG)}^{M'}\{ {\rm (CG)}^M, T^a\}]
\end{equation}
For shallow bound states, in the absence of any long-range interaction, the integral above can be computed using the
effective range expansion \cite{Bethe:1949yr,Bethe:1950jm} and is independent of the
details of the nuclear potential to leading order. 
As shown in Appendix \ref{app:nucl} the resultant short distance cross section in the low velocity regime is,
\begin{equation}
(\sigma v_{\rm rel})_{aMM'}^{\rm mag}=\kappa^2\frac {2^8}{g_N^2} \sigma_0   \left(1- \frac{M_a^2}{E_B^2}\right)^{\frac 3 2}\left(\frac {E_B} {M}\right)^{\frac 3 2}(1-a_{\mathbf{r}}\gamma_{\mathbf{r}'})^2  |C_J^{a M M'}|^2\,,\quad\quad \sigma_0\equiv\frac{\pi \alpha}{M^2}\,.
\label{xsec:magnetic}
\end{equation}
where $a_{\mathbf{r}}$ and $1/\gamma_{\mathbf{r}'}$ are the scattering
lengths associated to the initial- and final-state
channels.\footnote{We correct here an erroneous kinematical factor
  in  \cite{Redi:2018muu}.}  The same result can be derived using
nuclear effective field theory techniques, see \cite{Kaplan:2005es}.\footnote{Deuterium could also form by pion emission, if kinematically allowed.
This requires large binding energies, $E_B > M_\pi$. The leading interaction with pions is derivative,
\begin{equation}
c_N \frac {\partial_\mu \Pi^a}f \bar{N} \gamma^\mu \gamma^5 \ T^a N \Longrightarrow  \frac {\partial_i \Pi^a}f \bar{N} \sigma^i  T^a N
\end{equation}
where in the second step we took the non-relativistic limit.

This interaction has the same structure as magnetic dipoles so the same selection rules apply:
$\Delta S=1$, $\Delta L=0$, $\Delta I=0$. Hence the nuclear
cross section for pion emission is naively similar to that for a
magnetic transition (\ref{xsec:magnetic}), but
with $\sigma_0\to 1/f^2$.  For pion emission however $\sqrt{M E_B}>
M_\pi$, so the effective theory of nucleons cannot be used and a full
strongly-coupled computation is required. If allowed this process is
expected to give the largest production cross section.}
\paragraph{Electric transitions.~\\}
If the bound states are minimally coupled to the mediator an electric
dipole interaction is unavoidable. Computation of the cross section
proceeds as in the magnetic case and the analogous formula for the formation of an s-wave bound state through an electric dipole interaction reads \cite{Mitridate:2017izz},
\be
\begin{aligned}\label{eq:eleQM}
&(\sigma v_{\rm rel})_a^{\rm el}=(2S+1) \frac{2^4}{3 g_N^2}\frac{\alpha k}{M^2}\left(1-\frac{k^2}{3\omega^2}\right)\times \\
&\times \Bigg|\int r^2drR_{p,ij}\Bigg(\frac{1}{2}\left(T_{i'i}^a\delta_{jj'}-\overline{T}_{j'j}^{a}\delta_{ii'}\right) \partial_r
+i\,\frac{\alpha M}{2} T_{i'i}^b \overline{T}_{j'j}^cf^{abc}\frac {e^{-M_c r}(2+2 M_c r)-e^{-M_b r}(2+2 M_b r) }{r^2(M_b^2-M_c^2)}\Bigg)R^*_{B,j'i'}\Bigg|^2\,.
\end{aligned}
\ee
where the second term in the integral is associated to non-abelian
interactions.

As for magnetic transitions explicit formulae can be written in the
symmetric limit for shallow bound states, in the absence of
long-distance interactions.
Using the results in Appendix \ref{app:nucl} one finds,
\begin{equation}
(\sigma v_{\rm rel})_{aMM'}^{\rm el}=\frac {2 S+1}{g_N^2}  \frac {2^6}3 \sigma_0 v_{\rm rel}^2 \sqrt{1- \frac{M_a^2}{E_B^2}} \sqrt{\frac {M} {E_B}} \left(1+\frac{M_a^2}{2E_B^2}\right)|C_{J}^{a M M'}|^2
\label{xsec:electric}
\end{equation}
where we neglect the non-abelian contribution as it is typically suppressed for strongly-coupled bound states.
Note that in the electric case the cross section is not enhanced by a large scattering length in the initial state. 

\bigskip
Equations
\eqref{eq:magQM} and \eqref{eq:eleQM} are completely general and can be used to compute the
bound-state formation cross section for arbitrary potential and gauge
groups. 
With these tools at hand we proceed to study the
cosmology and astrophysics of DM bound-state formation, within the context
of two examples that highlight the relevant physics processes in the
strongly- and weakly-coupled cases.

\section{Strongly-coupled model: Weak-triplet dark baryon}
\label{sec:darkSU2}

The simplest model of nuclear DM is an $SU(3)_D$ gauge
theory with fermions in the triplet representation of SU(2)$_L$
\cite{Antipin:2015xia}; DM is the neutral component of an isospin-triplet (dark)
baryon $V$. If it is a thermal relic its abundance is determined by its pair-annihilation into dark pions.
Estimates of the rate for this process, substantiated by real QCD data,
indicate a thermal mass around 100 TeV. For smaller masses the correct
relic abundance can be realized instead through an asymmetry in the dark
sector completely analogous to that in the visible sector, with the DM made
up of the lightest neutral baryon. We will assume here that DM is
asymmetric, in which case the mass can be taken as a free parameter.
We emphasize however that the signals discussed here exist even in symmetric scenarios.

The dark nuclear forces will likely give rise to larger nuclei too,
with the lightest nucleus in each baryonic sector being absolutely stable
if $M_{\rm nucleus}< n M_{\rm nucleon}$ for a nucleus composed
of $n$ constituent nucleons, due to dark baryon number conservation. We will focus
here on production of the nucleus with baryon number 2, or `dark
deuterium', $D$. Its isotopes can be decomposed by their weak-isospin representations as,
\begin{table}[h!]
\begin{center}
\begin{tabular}{c|cccc}
\hbox{Isotope}& $\mathbf{r}$ & $S$    &$\lambda$   \\ \hline  
$D_1$ &$1$ &  0&2     \\ 
$D_3$ & 3 & 1  &1 \\ 
$D_5$ & 5 & 0  & -1   \\  
\end{tabular}
\quad \quad \quad
\caption{Dark deuterium isotopes made up of SU(2)$_L$-triplet
  dark baryons $V$. 
  $\mathbf{r}$ denotes the representation, labeled by its size, $S$ is 
spin, and $\alpha_{\rm eff}=\lambda \alpha $ is the effective gauge coupling. 
$D_{1,3,5}$ are branches of different SU(3) flavor  representations that are thus split by strong and electroweak
  interaction.  Smaller representations are expected to be more tightly bound.}
\label{table:boundstates}
\end{center}
\end{table}\\
Like for the wino, electroweak loop corrections will give rise to a
  small mass splitting between the charged and neutral components of
  the dark baryon $V$, $\Delta  \sim 165$ MeV for $M\sim 3$ TeV.  Although this splitting plays no part in the DM cosmology at
  early times, it becomes important at late times as the density of
  the charged component becomes negligible.

\subsection{Cosmological abundance}

The cosmological abundance of dark deuterium was estimated in
\cite{Redi:2018muu}, using the assumption of  factorization for the Sommerfeld-enhanced cross section.  
Here we improve on our previous computation by evaluating the cross sections numerically, 
including the full nuclear and electroweak potentials in our determination of the initial-state wavefunction. 

\begin{figure}[t]
\centering
\includegraphics[width=.45\textwidth]{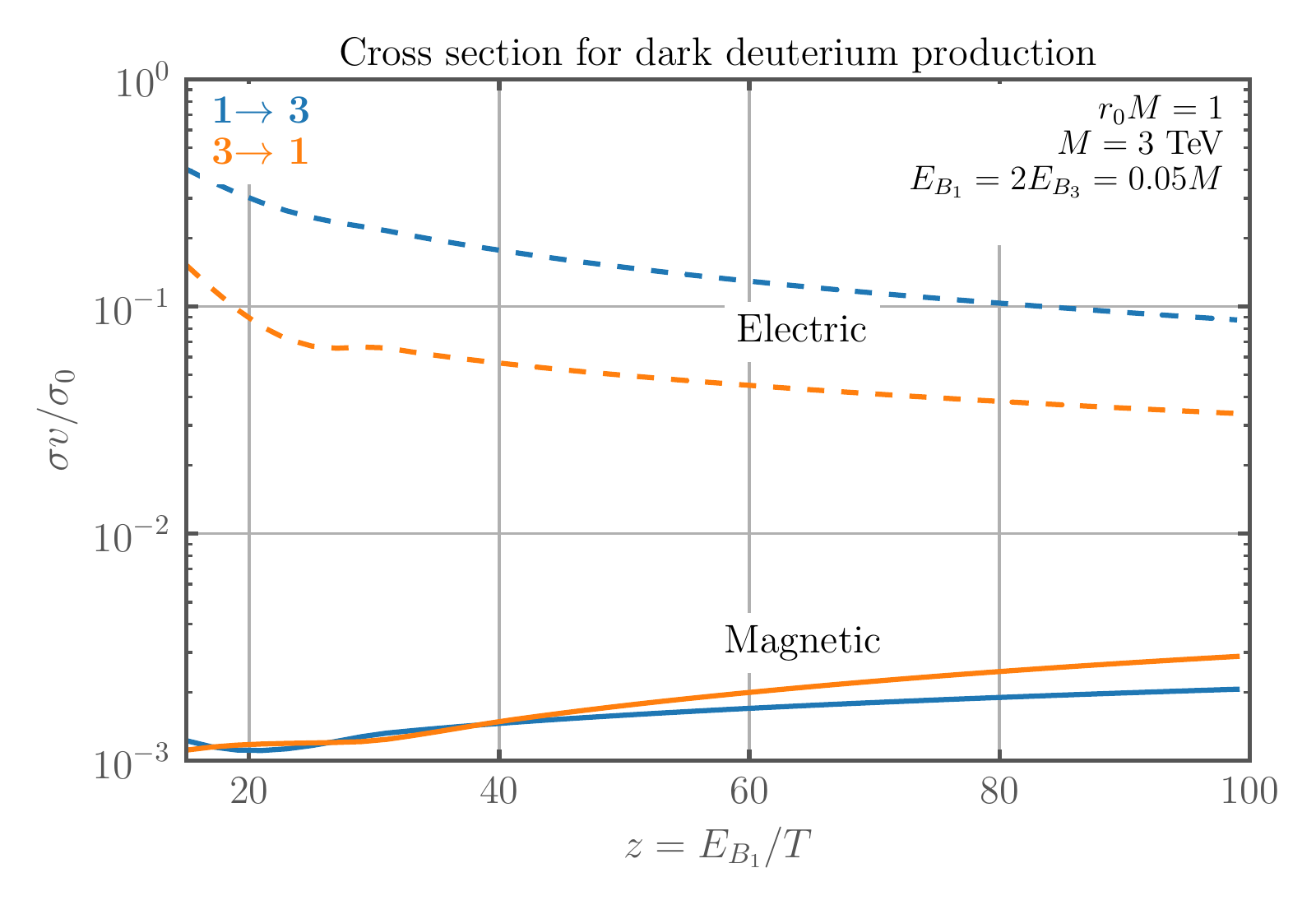}\quad
\includegraphics[width=.45\textwidth]{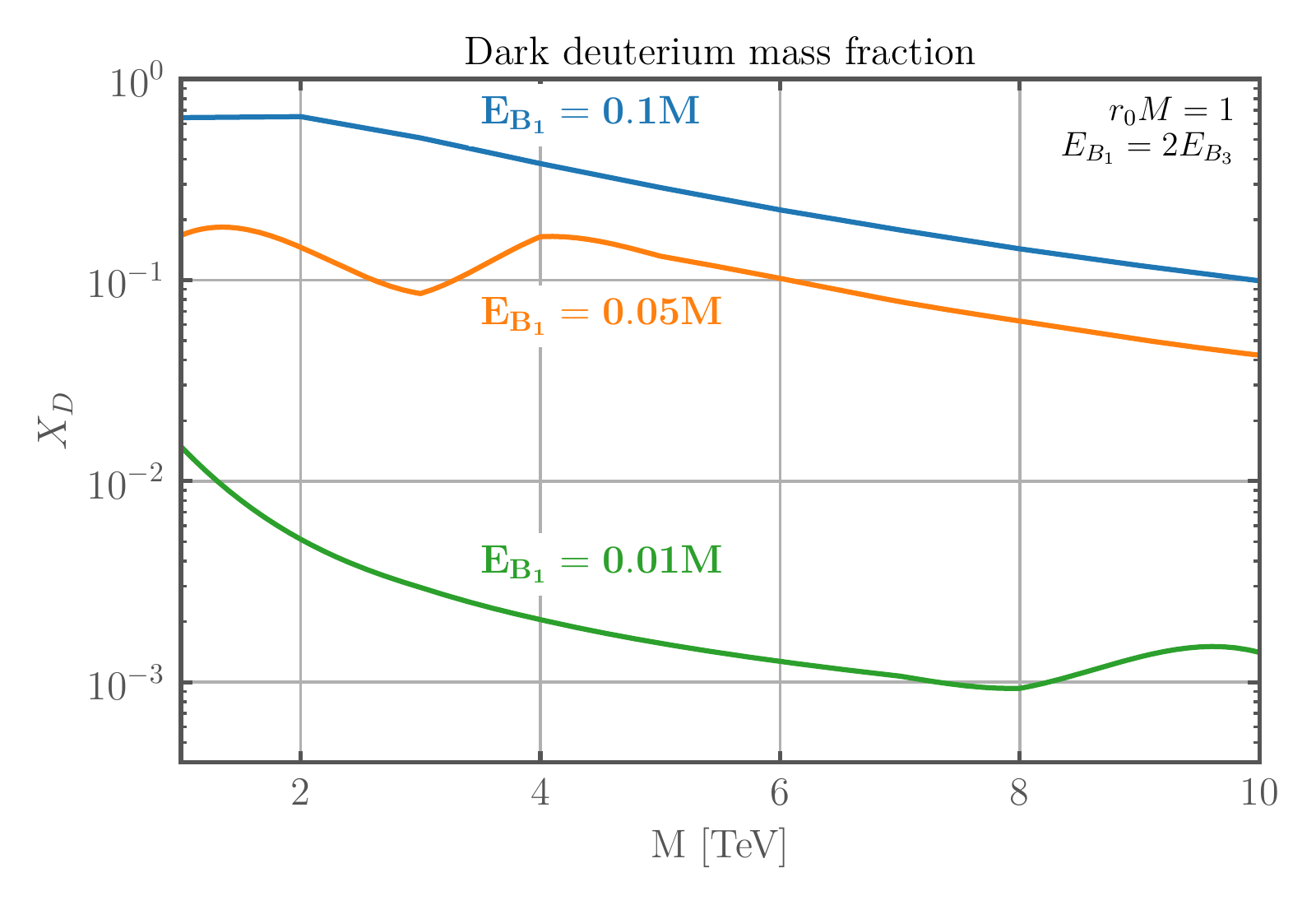}
\caption{\label{fig:dark-deuterium} Left panel: Dark deuterium
  production cross sections (rescaled) in different channels as a function of $z=E_B/T$.
Right panel: dark deuterium mass fraction $X_D$ as a function of the nucleon mass $M$ for various choices of parameters.
}
\end{figure}

We  assume for simplicity that DM is asymmetric with the dark baryon $V$ as the sole component, and that
only the singlet and triplet isotopes of dark deuterium $D_{1,3}$ are bound.  The DM abundance can be
computed by solving the following Boltzmann equation for the dark-deuteron
formation process $V + V \to D + X$, where $X$ stands for an
electroweak gauge boson $\gamma/Z/W$ in equilibrium with the SM
thermal bath, 
\begin{equation}\label{eq:2to2}
\dot n_D+ 3H n_D= \langle (\sigma v_{\rm rel})^{\rm eff}\rangle\left[ n_{V}^2- \frac {(n_{V}^{\rm eq})^2}{n_D^{\rm eq}}n_D\right]\,,
\end{equation}
where  $n_V$ and $n_D$ are the densities of dark baryons and dark
  deuterium, respectively, and $\langle (\sigma v)^{\rm eff}\rangle$ is the inclusive thermal
cross section for dark deuterium production. In the presence of several isotopes the equation
above describes the total deuterium abundance with the identifications
\begin{equation}\label{effective-xsec}
(\sigma v_{\rm rel})^{\rm eff} = \sum_i (\sigma v_{\rm rel})_{i}\,,~~~~~~~~~~~g_D^{\rm eff}(T)= \sum_i g_{D^i} \exp{\left[- \frac{E_{B_1}-E_{B_i}}{T}\right]}\,.
\end{equation}
For each value of the mass we compute the cross section numerically as a function of velocity and then integrate the Boltzmann equation above, see \cite{Redi:2018muu} for more details.
The dark deuterium production cross section due to electric and
  magnetic transitions, as well as the dark deuterium mass fraction
  for various choices of parameters, are shown in
  Fig.~\ref{fig:dark-deuterium}. A significant fraction of dark
deuterium is produced only for binding energies larger than in ordinary nuclear physics. 
For small binding energies the abundance is, moreover, smaller than estimated in \cite{Redi:2018muu}.

\subsection{Indirect Detection}
Since DM today consists of  the neutral component of the dark
baryon triplet, only the two-particle DM subsector with zero charge will be relevant for indirect detection.
The radial wavefunction is described by a two-component vector,
\begin{equation}
R=\left( R_{+}\,,R_{0}\right)\,,
\label{radial-W}
\end{equation}
corresponding to the $V^+V^-$ and $V^0V^0$ two-particle states,
respectively. The s-wave initial state has spin 0 while 
p-wave state has spin 1. In both cases the static potential has the form,
\begin{equation}
V_{Q=0}^{S=0}=  \left( \begin{array}{cc} 2\Delta - A & -\sqrt{2} B \\
                         -\sqrt{2} B & 0 \end{array}\right) + V_N^{\bf
                     1}(r)  \left( \begin{array}{cc} \frac{2}{3} &
                                                                    \frac{\sqrt{2}}{3} \\ \frac{\sqrt{2}}{3} & \frac{1}{3} \end{array}\right)  + V_N^{\bf 5}(r)  \left( \begin{array}{cc} \frac{1}{3} & -\frac{\sqrt{2}}{3} \\ -\frac{\sqrt{2}}{3} & \frac{2}{3} \end{array}\right) \,,
\label{eq:wino-potential}
\end{equation}
where $A = (\alpha_{\rm em} + \alpha_2 c_W^2 e^{-M_Z r})/r$, $B =
\alpha_2  e^{-M_W r}/r$ with $c_W=\cos\theta_W$ are the electroweak
contributions, and 
$\Delta=166\,\mathrm{MeV}$ is charged-neutral mass splitting. 
$V_N^{\bf 1,5}$ are the nuclear potential in the singlet and quintuplet isospin channel.
In our numerical study we approximate the nuclear potentials with a spherical well with 
a depth chosen to support bound states of given binding energies, see Appendix \ref{app:nucl}.  

Since electric and magnetic interactions both violate isospin by one
unit, isospin selection rules allow us to form an isospin-triplet
nuclear bound state (spin 1) from an isospin singlet or
quintuplet in the initial state. 
The electrically-neutral component is supported by the potential,
\begin{equation}
V_{Q=0}^{S=1}= 2 \Delta - A +V_N^{\bf{3}}(r)
\end{equation}
The triplet bound state is approximately isospin symmetric for the choice of parameters we consider.
In isospin space the wave function is determined by Clebsch-Gordan coefficients $|1 0 \rangle= \frac 1 {\sqrt{2}} | V^+ V^-\rangle - \frac 1 {\sqrt{2}} | V^-V^+\rangle$;
the radial wavefunction $R_B(r)$ can be conveniently chosen to be real (see Appendix \ref{app:nucl} for full form of bound-state wavefunction).

In the basis of Eq.~\eqref{eq:magQM} the initial- and final-state wavefunctions have the form:
\begin{equation}
\begin{split}
&R_{ij}^l(r) = {\rm Diag}[R_{+}^{(\ell)}/\sqrt{2}\,,R_0^{(\ell)}\,,R_+^{(\ell)}/\sqrt{2}]\\
&R_{ij}^{B_0}(r)= {\rm Diag}[1/\sqrt{2}\,,0\,,-1/\sqrt{2}]R_B  
\label{eq:Rijnotation}
\end{split}
\quad \quad R_{ij}^{B_+}(r)= \left( \begin{array}{ccc} 0 & 1  & 0 \\ 0   & 0 &  -1 \\ 0 & 0 & 0  \end{array}\right)\frac{R_B}{\sqrt{2}}
\end{equation}
where $R^{(\ell)}$ for $\ell=s,p$ are the radial wavefunctions for
the initial-state s- and p-wave states, while $R_B(r) $ is 
the radial wavefunction of the isospin 1 bound state. 

\subsubsection{Photon lines}
Let us now discuss astrophysical signals from photon emission on
dark deuterium formation, and their measurement, either in the
galactic center or in more distant dwarf galaxies.  Since DM
today is highly nonrelativistic ($\beta=10^{-3}$ and $\beta=10^{-4}$, respectively) the photons emitted are
monochromatic up to energy resolution effects, with $E_\gamma= E_B$,
the binding energy of the final state.

Recall from Table \ref{table:boundstates} and the subsequent
discussion, that selection rules allow for the formation of the neutral
isospin-triplet spin-1 dark deuteron $D^0_{\mathbf{3}}$. This state
can be formed by a magnetic transition from an $s$-wave initial
state, or equivalently by an electric transition from a $p$-wave spin-1 initial
state. From eqs. (\ref{eq:magQM},\ref{eq:eleQM}) ($T^a=-\bar{T}^a=J_3$ and $g_N=2$) the cross sections for these processes are:
\begin{equation}
(\sigma v_{\rm rel})^{\rm mag}_{D^0_{\mathbf{3}}\gamma}= 8 \kappa^2\,
\alpha_{\rm em} \frac{ E_B^3}{M^2}\times
\Bigg|\int r^2dr R^{(s)}_{+} R_B\Bigg|^2\,,
\end{equation}
\begin{equation}
(\sigma v_{\rm rel})^{\rm el}_{D^0_{\mathbf{3}}\gamma}=\frac 8 3
\alpha_{\rm em} \frac{E_B}{M^2}\times
\Bigg|\int r^2dr R^{(p)}_{+}  \partial_r R_B \Bigg|^2\,,
\end{equation}
where we neglected the electroweak non-abelian contribution in the
electric transition as it is suppressed for small nuclear bound states.


\begin{figure}[t]
\centering
\includegraphics[width=0.48\textwidth]{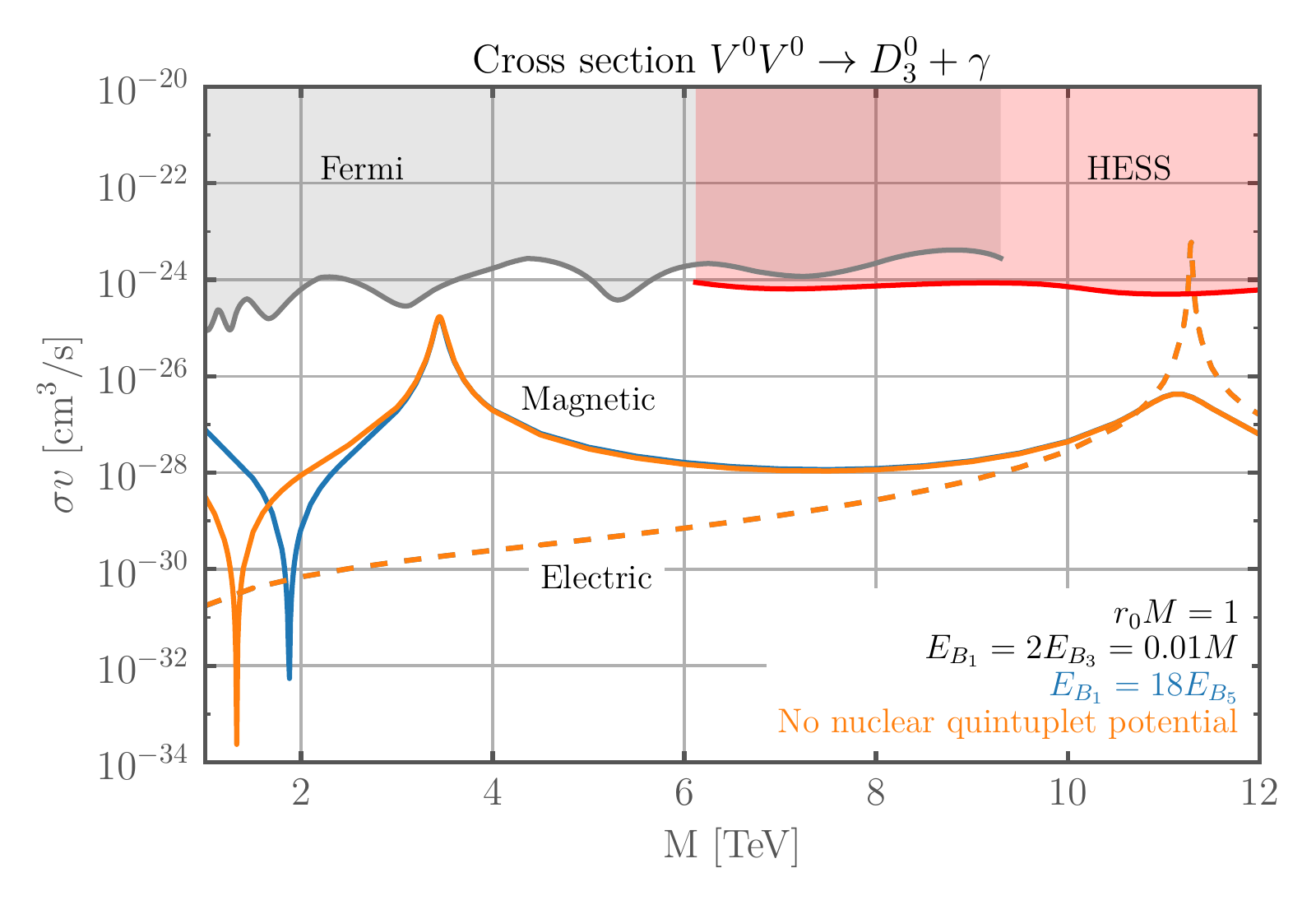}~
\includegraphics[width=0.48\textwidth]{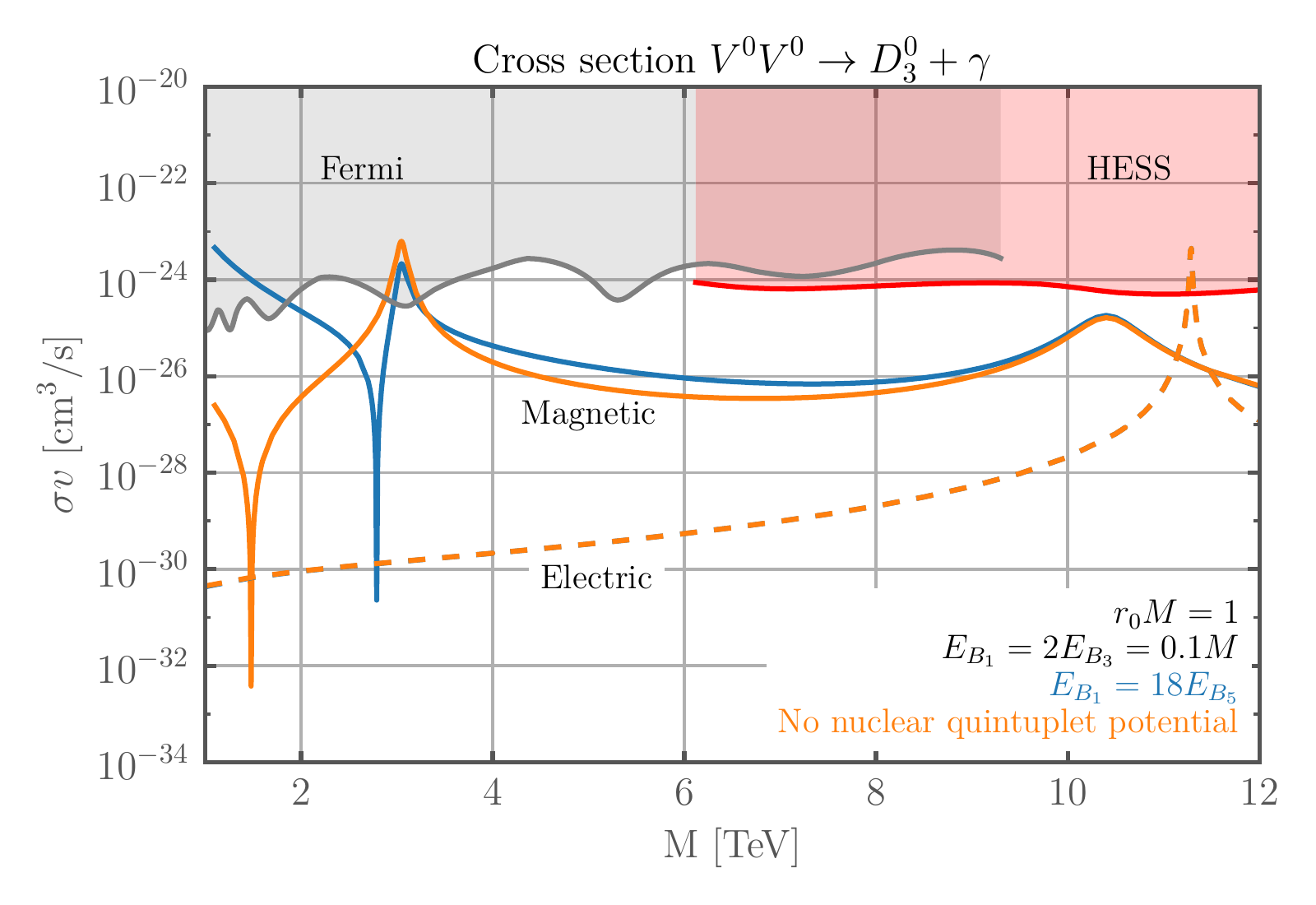}
\caption{\label{fig:xsec3g}
Cross section for monochromatic photon production for the process $V^0 V^0 \to D_3^0 +\gamma$ via the electric and magnetic transitions for different choices of nuclear potentials.
The gray (red) region is the exclusion due to $\gamma$-ray lines from
FERMI \cite{FERMI-LAT} (HESS \cite{HESSDW}) from our galactic centre,
assuming an NFW DM profile. We take $\beta=10^{-3}$.}
\end{figure}

In Fig.~\ref{fig:xsec3g} we present the cross section for bound state
formation by monochromatic photon emission from the galactic centre
(for bounds from dwarf spheroidal galaxies see \cite{Lefranc:2016fgn}).
Unlike photons produced from DM annihilation, the energy $E_B$ of these
photons is independent of the DM mass $M$ and
so all experimental bounds for bound-state formation must be rescaled according to
\begin{equation}
\langle \sigma v_{\rm rel}\rangle_{D\gamma} < 2 \left( \frac {M}{E_B}\right)^2   \langle \sigma v_{\rm rel} \rangle_{\gamma \gamma}\Big|_{M_{\rm DM}=E_B}\,.
\label{eq:rescaling}
\end{equation}
This rescaling takes into account  the reduced DM numerical density compared to annihilation into gauge bosons of DM with $M_{\rm DM}=E_B$
and that a single gauge boson is emitted in bound state formation. 
We see in the plots some generic features discussed in Section
\ref{sec:factorization}.  The peaks of the cross section curves are
related to those in Sommerfeld-enhanced DM annihilation processes, but
for the magnetic transition are shifted to higher masses.  The
magnetic channel also features dips.  These can be understood as the
effect of negative interference. As shown in Appendix
\ref{app:Factorized} the short-distance nuclear annihilation matrix
has non-zero off-diagonal components, and electroweak SE allows $V^0
V^0 \to D^0_3 \gamma$ and $V^++ V^- \to D^0_3 \gamma$ to interfere
destructively.  Indeed the analogous effect is also realized for
annihilating DM when the reactions proceed through several channels, see \cite{Chun:2012yt,Cirelli:2007xd}.

Finally, as mentioned in \cite{Mahbubani:2019pij} $D_3^0$ is likely not the lightest dark-deuterium isotope. As a consequence a second monochromatic photon  
could be emitted in transition to the ground state $D_1^0$, with energy
$E_{\gamma'}=E_{B_1}-E_{B_3}$. If this was the only decay channel
kinematically allowed to the dark deuterium triplet state, its rate
would be
identical 
to that of the primary photon emission, giving rise to a smoking-gun signature of bound state formation. 

Current experiments are already sensitive to formation of dark
deuterium in the neighborhood of the peak of the of the cross section,
assuming an NFW DM profile. 
Future experiments such as HERD \cite{Fusco:2019zna} will improve the
bounds and yield a better energy resolution for the lines.

\subsubsection{Diffuse photons from $W/Z$ emission}

\begin{figure}[t]
\centering
\includegraphics[width=0.6\textwidth]{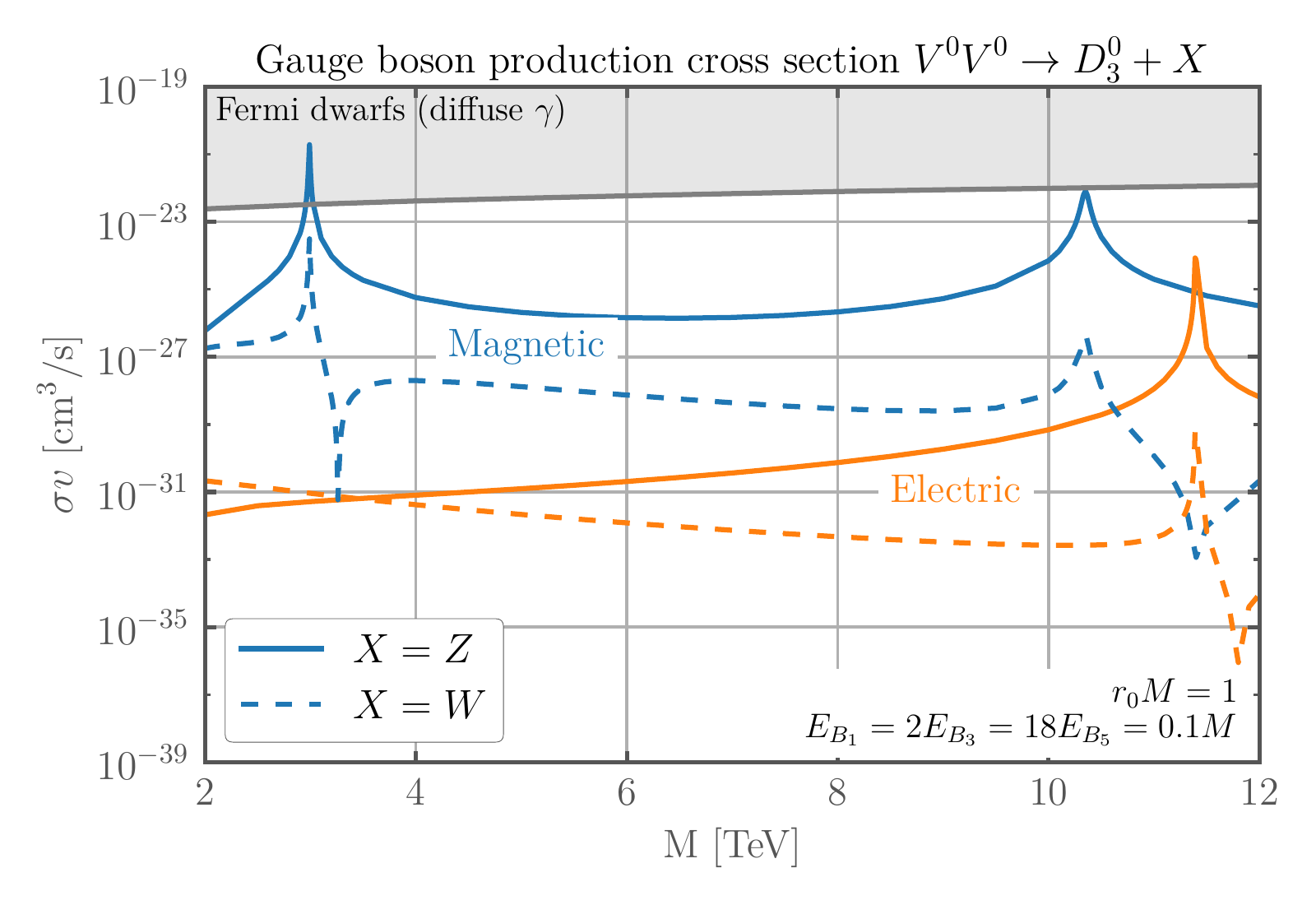}
\caption{\label{fig:tripletWZ}Cross sections for $W$ and $Z$ emission
  via formation of a dark-deuterium bound state, in electric and magnetic channels.
The resultant $\gamma$ ray spectrum is constrained by diffuse
$\gamma-$ray searches in dwarf spheroidal galaxies by FERMI \cite{Ackermann:2015zua}.}
\end{figure}

For binding energies larger than $M_{W,Z}$ bound-state production
could also proceed through emission of $W$ and $Z$ gauge bosons. The cross section for these
processes can be simply obtained from Eq. (\ref{eq:magQM}),
(\ref{eq:eleQM}), with the use of the relevant couplings and
generators ($T^a=-\bar{T}^a=J_3$ for $Z$ emission, $T^a=-\bar{T}^a=J_+$ for $W$).
One finds,
\begin{equation}
\begin{split}
&(\sigma v_{\rm rel})^{\rm mag}_{D^+_{\mathbf{3}}W^-}=4 \kappa^2\, \alpha_2\frac{ E_B^3}{M^2}\left(1- \frac {M_W^2}{E_B^2}\right)^{\frac 3 2} \times
\Bigg|\int r^2dr \left(R_{0}^{(s)}-\frac{R_{+}^{(s)}}{\sqrt{2}}\right) R_B\Bigg|^2 \\
&(\sigma v_{\rm rel})^{\rm el}_{D^+_{\mathbf{3}}W^-}= \frac 4 3
\alpha_2 \frac{E_B}{M^2}\sqrt{1- \frac {M_W^2}{E_B^2}}\left(1+\frac
  {M_W^2} {2E_{B_{\mathbf{3}}}^2}\right) \times
\Bigg|\int r^2dr \left(R_{0}^{(p)}- \frac{R_{+}^{(p)}}{\sqrt{2}}\right) \frac{\partial}{\partial r} R_B \Bigg|^2 \\
&(\sigma v_{\rm rel})^{\rm mag}_{D^0_{\mathbf{3}}Z}= 8 \kappa^2\, \alpha_2 c_W^2 \frac{ E_B^3}{M^2}\left(1- \frac {M_Z^2}{E_B^2}\right)^{\frac 3 2} \times
\Bigg|\int r^2dr R_{+}^{(s)} R_B\Bigg|^2\\
& (\sigma v_{\rm rel})^{el}_{D^0_{\mathbf{3}}Z}=\frac 8 3\alpha_2 c_W^2  \frac{E_B}{M^2}\sqrt{1- \frac {M_Z^2}{E_B^2}}\left(1+\frac {M_Z^2} {2E_{B_{\mathbf{3}}}^2}\right)\times
\Bigg|\int r^2dr R_{+}^{(p)} \partial_r R_B \Bigg|^2
\end{split}
\end{equation}
The inclusive $W$ cross section is twice the value above.

In Fig.~\ref{fig:tripletWZ} we compare the cross section for $W$ and
$Z$ emission on production of a dark-deuterium bound state, with the experimental bound on
diffuse photon emission from dwarf spheroidal galaxies by FERMI \cite{Ackermann:2015zua}.
Close to the peaks the magnetic cross section can give rise to a signal
that within the sensitivity of the experiment; this behaviour is consistent with expectations from \cite{Chen:2013bi}.

Bound state formation is also constrained by the CMB that places a rather model independent bound on the energy injected 
in the photon plasma around recombination. As discussed in \cite{Mahbubani:2019pij} the bound on the total cross-section for bound state formation takes the form,
\begin{equation}
\langle \sigma v_{\rm rel}\rangle_{\rm CMB}< \frac {8.2 \times 10^{-28}\, {\rm cm^3}{s^{-1}}}{f_{\rm eff}}\times  \frac {M^2}{E_B^2}\times  \frac {E_B}{\rm GeV}\,,
\end{equation}
where $f_{\rm eff}\approx 0.5$ depends mildly on the decay channel. This bound is competitive with the one from dwarf spheroidal galaxies reported in Fig. 6 for masses in the TeV range,
see  \cite{Bringmann:2016din,Cirelli:2016rnw,Baldes:2017gzu}.

\section{Weakly-coupled model: Weak-quintuplet Minimal
  DM}
\label{sec:mdm}

The tools developed in this paper can also be used to study signals of
bound-state formation in weakly-coupled models. In this section we
apply these techniques to minimal DM, in the form of a
fermionic SU(2)$_L$ quintuplet with zero hypercharge \cite{Cirelli:2005uq}. 
Including
the effects of bound-state formation in the computation of the minimal
DM
relic density gave rise to a significant increase in thermal relic mass, from 9.5 TeV
\cite{Cirelli:2007xd} to  14$\pm$1 TeV \cite{Mitridate:2017izz}.
We expect existing indirect-detection bounds on minimal DM to be
similarly modified on correctly accounting for bound-state formation,
both in the annihilation channel, which results in emission of gauge
bosons with energies equal to the DM mass, and the bound-state
formation channel, where the emitted gauge bosons have energy equal to
the binding energy of the final state.  We update the existing
phenomenological constraints in this section.

\subsection{Annihilation}
\begin{figure}[t]
\centering
\includegraphics[width=.49\textwidth]{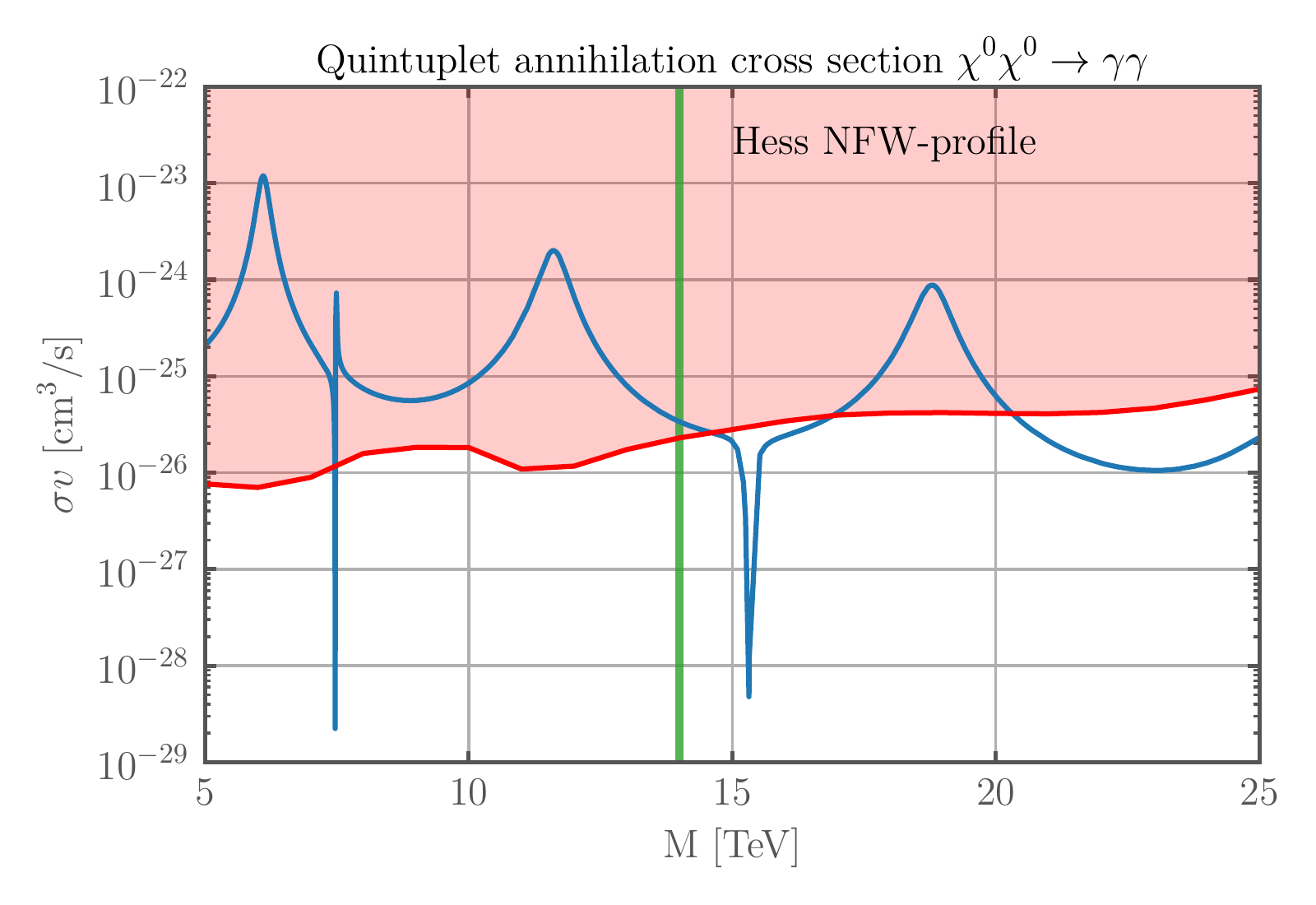}~
\includegraphics[width=.49\textwidth]{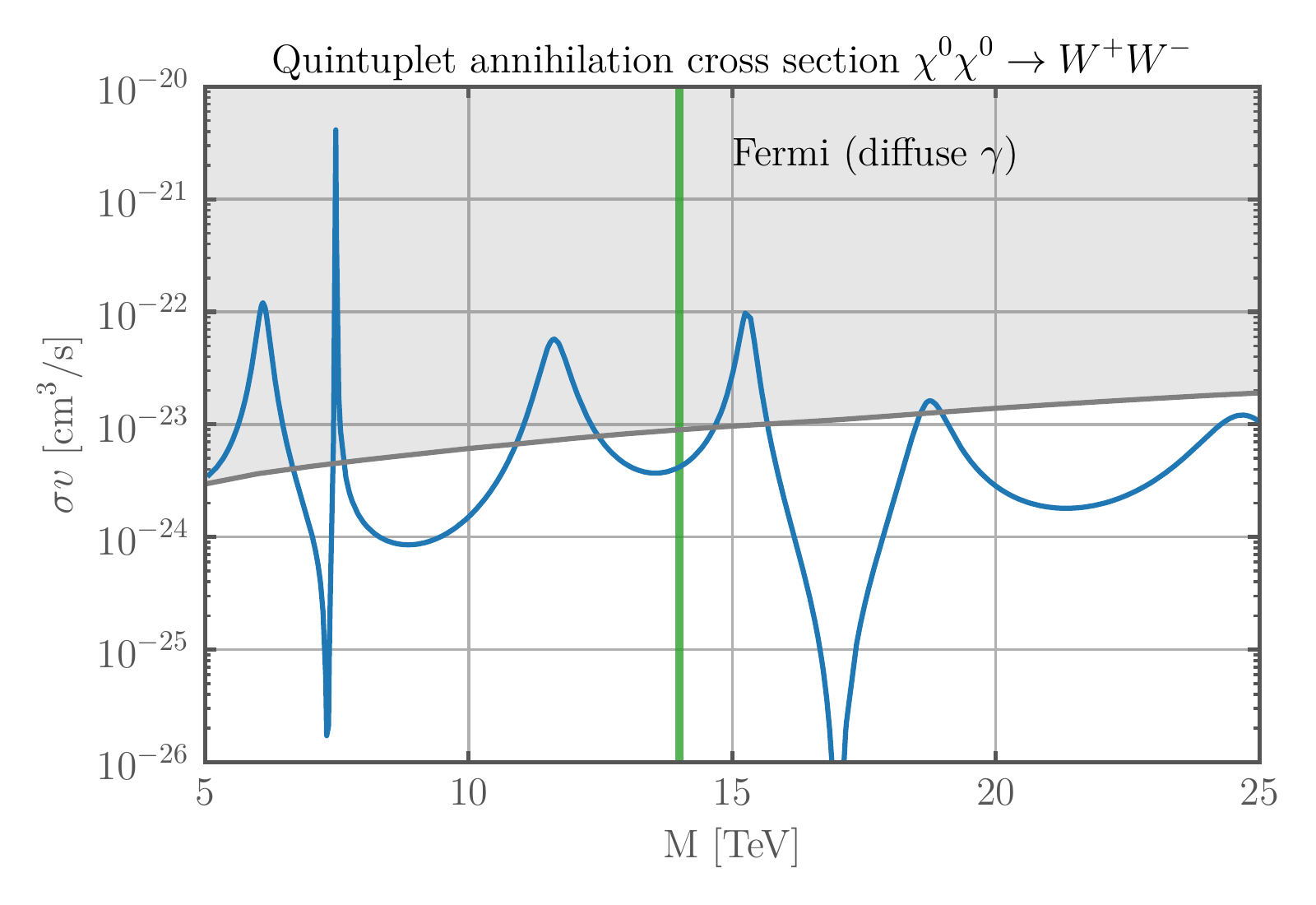}~
\caption{\label{fig:x-sec5} Left panel: Annihilation cross section of
  minimal DM SU(2)$_L$-quintuplet into photons for $\beta=10^{-3}$. The experimental bound are extracted from Galactic center HESS data
assuming an NFW profile \cite{HESSGC}. Right panel: Annihilation into WW
leading to a diffuse photons signal. We use here the same bounds as in \cite{Cirelli:2015bda} from the galactic center.}
\label{fig:DF}
\end{figure}
The annihilation rate for the neutral component of the quintuplet,
$\chi^0$, is strongly modified by inclusion of SE in the initial
state.  Although a detailed computation can be found in
\cite{Cirelli:2015bda}, we extend their analysis to the larger range
of DM
masses that are thought consistent with updated relic density
considerations.  As in the previous example, since DM is a Majorana
fermion the relevant sector has spin-0 and is electrically neutral, $(Q=0, S=0)$. It encapsulates mixing between the following
two-particle states:
$\{\chi^{++}\chi^{--},\chi^{+}\chi^{-},\chi^0\chi^0\}$, and the
associated electroweak potential in this basis is
 \begin{equation}
V_{Q=0}^{S=0}= \left( \begin{array}{ccc} 8\Delta - 4 A & -2 B & 0 \\  -2 B & 2 \Delta-A & -3 \sqrt{2} B \\0 & -3 \sqrt{2} B & 0  \end{array}\right) \,,
\label{5potential}
\end{equation}
where the coefficients $A$ and $B$ were defined below Eq.~\eqref{eq:wino-potential}.

For simplicity we only consider the final states ($\gamma\gamma,WW$), with annihilation matrices
\begin{equation}
\Gamma_{Q=0}^{S=0}\Big|_{\gamma\gamma}= \frac {\pi \alpha_{\rm em}^2}{ M^2}  \left( \begin{array}{ccc} 16 & 4 & 0 \\  4 & 1 & 0  \\ 0 & 0 & 0  \end{array}\right) \,, \quad \quad
\Gamma_{Q=0}^{S=0}\Big|_{WW}= \frac {\pi \alpha^2_{2}}{2 M^2}  \left( \begin{array}{ccc} 36 & 18 & 6 \sqrt{2} \\  18 & 27 & 14 \sqrt{2}  \\ 6 \sqrt{2} & 14 \sqrt{2} & 18  \end{array}\right) \,.
\label{5annihilationrates}
\end{equation}
We compute the total cross section using factorization, and
  including full isospin-breaking effects in the SE as detailed in
Appendix \ref{app:Factorized}.
The cross sections for annihilation of DM into pairs of
(monochromatic) photons of energy $M$, as well as to $W$ bosons are
displayed in Fig.~\ref{fig:x-sec5}. Superposed on these plots are the
bounds on monochromatic photons from the galactic center from HESS, as
extracted from the 10-year line search \cite{HESSGC}, and the NFW
limit from diffuse photons searches by Fermi, as extracted from Fig. 4
of Ref.~\cite{Cirelli:2015bda}. We see that for $M=14$ TeV where the thermal abundance of 
DM is nominally reproduced the signal is bordeline but still consistent with $\gamma-$ray constraints
even with the aggressive NFW profile. Different values of the mass might be however excluded.

\subsection{Bound-state formation}
In this section we consider the indirect signals of a minimal DM
quintuplet pair undergoing bound-state formation.  A preliminary study in
the SU(2)$_L$-symmetric approximation can be found in
\cite{Mitridate:2017izz}. Our approach allows us to compute the
relevant rates without relying on the approximate symmetric limit, but
instead taking into account the full potential including electroweak
splittings in the initial state

Bound state formation in this case has two potential paths to an
indirect-detection signature.  The first is due to the SM gauge boson
that is emitted as a by-product, and the second is directly from the
decays of the unstable bound state produced.  However, since the main
production channel is to a spin-1 bound state that subsequently
annihilates to the SM Higgs and fermions, no significant bounds are
obtained from this channel.

The leading transition mechanism for weakly-coupled bound states is
via the electric coupling to electroweak gauge
bosons. Recall that electric dipole interactions imply the selection
rules $\Delta S=0$, $\Delta L=1$. Hence s-wave bound states are formed
from a p-wave initial state while p-wave bound state are formed from s-wave
and d-wave initial states. Decomposing the DM initial state in
eigenstates of isospin
\begin{equation}
|\chi^0 \chi^0\rangle = \frac 1 {\sqrt 5} |\mathbf{1},0\rangle - \sqrt{\frac 2 7}  |\mathbf{5},0\rangle +  \sqrt{\frac {18} {35}}  |\mathbf{9},0\rangle.
\end{equation}
Given that in the dipole approximation $\Delta I=1$ it follows that
only the triplet bound state can be produced (the septuplet is not
bound). The bound states produced in the process are approximately coulombian
and can be easily computed and classified by their isospin $I$, spin
$S$, and angular momentum $(n,\ell)$ \cite{Mitridate:2017izz}.  For the critical mass $M_*=14$ TeV 
the binding energies of the relevant states are given by
\begin{equation}
\begin{tabular}{ccccc|c|cc|c}
\hbox{Name}& $I$ & $S$ & $n$ & $\ell$ &$ E_B(M_*)/{\rm GeV}$   & \hbox{Produced from}\\  
$1s_3$ & 3 & 1 & 1 & 0 &100  & $p_1$, $p_5$ \\ 
$2s_3$ & 3 & 1 & 2 & 0 &25  & $p_1,p_5$ \\
$2p_3$ & 3 & 0 & 2 & 1 &25  & $s_1,s_5$ \\
\end{tabular}
\label{5boundstates}
\end{equation}
In the notation of eq. (\ref{eq:eleQM}) the initial wavefunction reads,
\be
R_{ij} ={\rm Diag}\left[R_{++}^{(\ell)}/\sqrt{2}\,,R_{+}^{(\ell)}/\sqrt{2}\,, R_{0}^{(\ell)}  \,, R_{+}^{(\ell)}/\sqrt{2}\,, R_{++}^{(\ell)}/\sqrt{2}\right]\,,
\ee
where $R^{(\ell)}_{++},\,R^{(\ell)}_{+},\,R^{(\ell)}_0$ are the
  $\ell$-wave radial wavefunction describing the initial states $(\chi^{++}\chi^{--})$,\,$
  (\chi^{+}\chi^-)$,\,$(\chi^0\chi^0)$ respectively. The final states are approximately SU(2)-symmetric
  and their wavefunction is determined by group theory up to the radial wavefunction. One finds,
\begin{equation}
R_{ij}^{B_0}= {\rm Diag}\left[2 \,,-1\,,0  \,, 1\,,-2 \right]\frac{R_B(r)}{\sqrt{10}} \quad \quad R_{ij}^{B_+}= \left( \begin{array}{ccccc} 0 & 1  &0 & 0 & 0 \\ 
0   & 0 & -\sqrt{\frac 3 2} & 0 & 0 \\
0 & 0 & 0 &\sqrt{\frac 3 {2}} & 0  \\ 0 & 0 & 0  & 0 & -1
 \\ 0 & 0 & 0  & 0 & 0
\end{array}\right) \frac{R_B(r)}{\sqrt{5}}\,.
\label{eq:Rijnotation5}
\end{equation}

\subsubsection{Photon lines} 
From the above decomposition we can now compute bound-state formation with
the emission of a photon. The energy of the emitted photon is equal to
the binding energy of the bound state, which is fixed by the DM mass
and the choice of quantum numbers in the initial and final states.
For thermal relic mass for instance the binding energies, as read from
\eqref{5boundstates} are  25 and 100 GeV.

The cross section for the formation of an s-wave isospin-triplet (spin-1) bound state,
\begin{equation}
\begin{split}
(\sigma v_{\rm rel})^{\rm el}_{D^0_{\mathbf{3}}\gamma}&=
\frac{16}{3}\frac{\alpha_{\rm em} E_B}{M^2}\\
&\times \Bigg|\int r^2dr\left[\left(2\sqrt{\frac 2 5}  R^{(p)}_{++}-\sqrt{\frac 1 {10}} R^{(p)}_{+}\right) \frac {\partial}{\partial r}  + \alpha_2 M e^{-M_W r}  \left(\frac{{3 R^{(p)}_{0}}-2\sqrt{2} R^{(p)}_{+} -\sqrt{2}R^{(p)}_{++}}{2\sqrt{5}}\right)\right] R_B\Bigg|^2
\end{split}
\label{eq:quintupletD3e}
\end{equation}
and to use it we need to solve the p-wave Schroedinger equation with
the potential given in Eq.~\eqref{5potential}.

\begin{figure}[t]
\centering
\includegraphics[width=.49\textwidth]{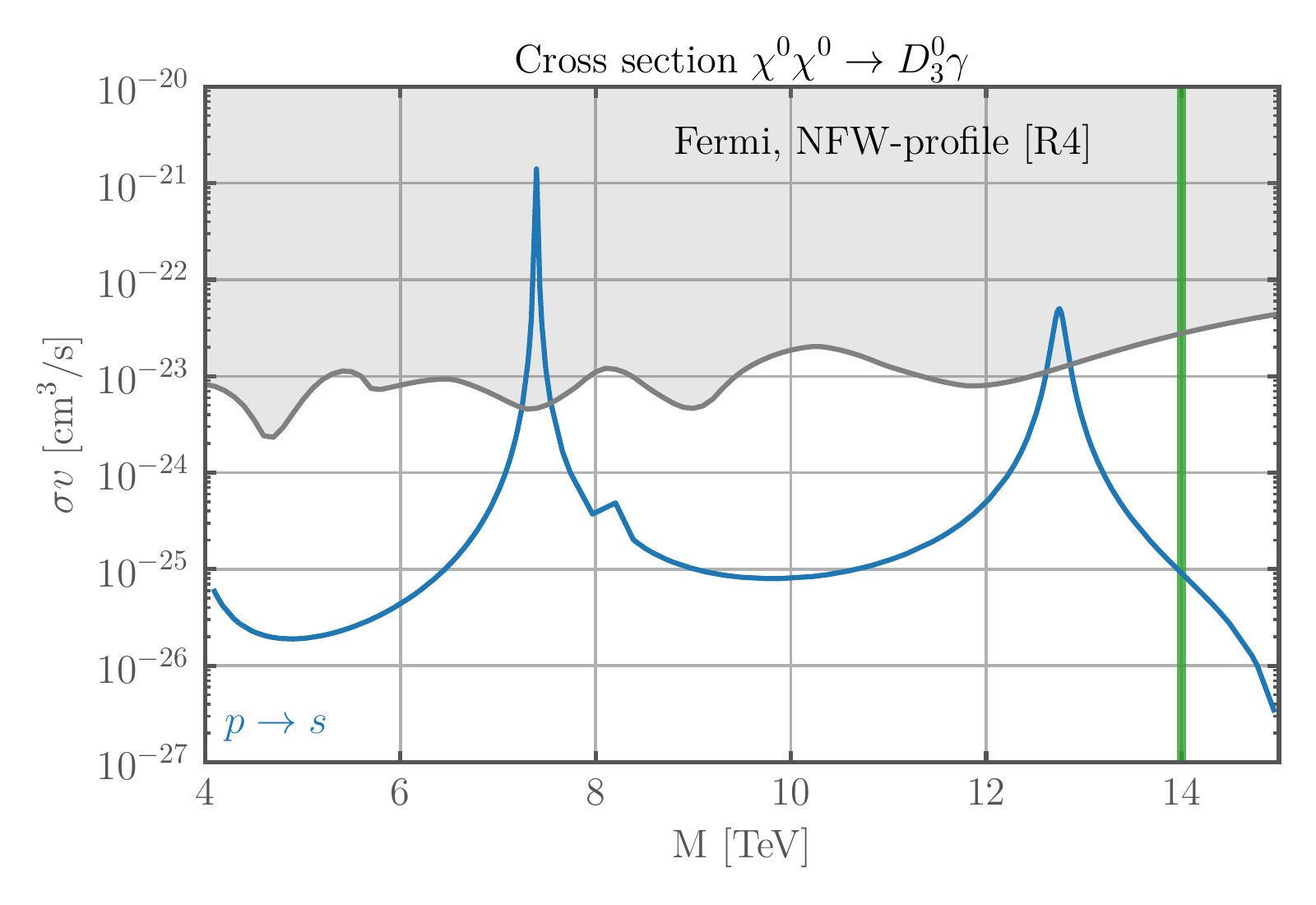} ~
\includegraphics[width=.49\textwidth]{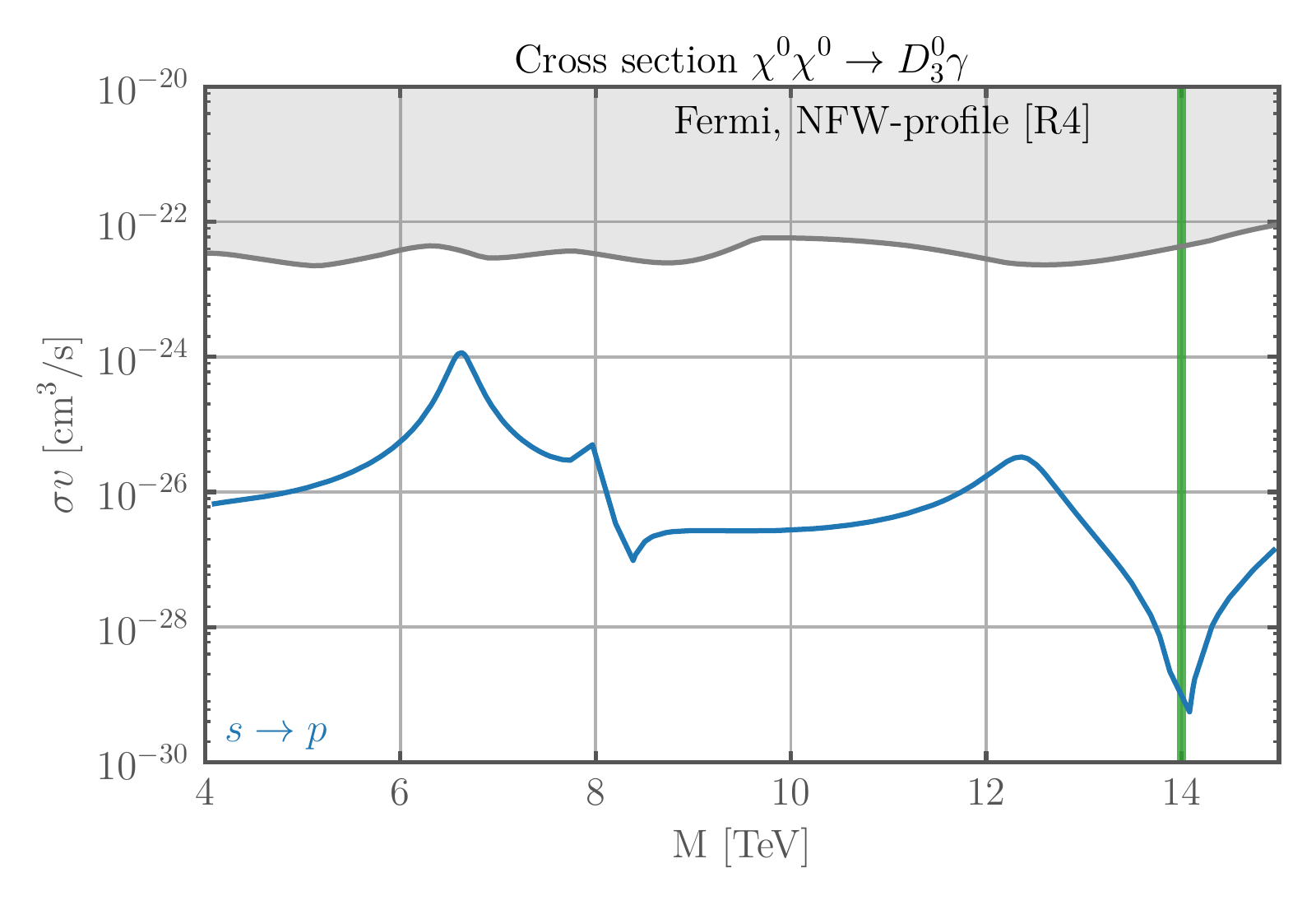}
\caption{\label{fig:BS5pletg} Cross section for bound state formation through emission of photons for s-wave (1$s_3$) and p-wave (2$p_3$) bound states produced from $\ell=1$ and $\ell=0$ partial waves respectively.  The strongest bounds are found for the production  of 1$s_3$ bound state.}
\label{fig:DF}
\end{figure}

For completeness the analogous cross section for the formation of a
p-wave isospin-triplet (spin-0) bound state is
\begin{equation}
\begin{split}
(\sigma v_{\rm rel})^{\rm el}_{D^0_{\mathbf{3}}\gamma}&=
\frac{16}{9}\frac{\alpha_{\rm em} E_B}{M^2}\\
&\times \Bigg|\int r^2dr \left[-\frac {\partial}{\partial r} \left(2\sqrt{\frac 2 5}  R^{(s)}_{++}-\sqrt{\frac 1 {10}} R^{(s)}_{+}\right)  + \alpha_2 M e^{-M_W r}   \left(\frac{3 R^{(s)}_{0}-2\sqrt{2} R^{(s)}_{+} -\sqrt{2}R^{(s)}_{++})}{2\sqrt{5}}\right)\right] R_B\Bigg|^2
\end{split}
\end{equation}
The monochromatic photon cross section due to bound-state formation in
these two channels are shown in Fig.~\ref{fig:BS5pletg}. For a
quintuplet mass of 14 TeV, bound state production does not yield a
significant constraint even in the larger $p\to s$ channel. We note however that the cross section is very sensitive 
to the mass motivating a precision study.

\subsubsection{Diffuse photons}
The binding energy of the  s-wave isospin-triplet bound state is
sufficiently large to also emit $W$ and $Z$ bosons in a sizeable
  interval around the mass where the thermal abundance is
  reproduced. The cross section for the latter can be obtained
by rescaling Eq.~(\ref{eq:quintupletD3e}) by the coupling strength and
kinematic emission factor for the $Z$ boson:
\begin{equation}
\begin{split}
(\sigma v_{\rm rel})^{\rm el}_{D^0_{\mathbf{3}}Z}&= \frac{16}{3}\alpha_2 c_W^2\frac{E_B}{M^2}\sqrt{1- \frac {M_Z^2}{E_B^2}}\left(1+\frac {M_Z^2} {2E_{B_{\mathbf{3}}}^2}\right) \times\\
&\times \Bigg|\int r^2dr \left[\left(2\sqrt{\frac 2 5}  R^{(p)}_{++}-\sqrt{\frac 1 {10}} R^{(p)}_{+}\right) \frac {\partial}{\partial r}  + \alpha_2 M e^{-M_W r}   \left(\frac{3 R^{(p)}_{0}-2\sqrt{2} R^{(p)}_{+} -\sqrt{2}R^{(p)}_{++})}{2\sqrt{5}}\right)\right] R_B\Bigg|^2
\end{split}
\end{equation}
For $W$ emission we have instead:
\begin{equation}
(\sigma v_{\rm rel})^{\rm el}_{D^+_{\mathbf{3}}W^-}= \frac{16}{3}\alpha_2 \frac{E_B}{M^2}\sqrt{1- \frac {M_W^2}{E_B^2}}\left(1+\frac {M_W^2} {2E_{B_{\mathbf{3}}}^2}\right) \times  \Bigg|\int r^2dr \left[\left( \frac{R^{(p)}_{++}}{\sqrt{10}}-\frac 1 2 \sqrt{\frac 5 2} R^{(p)}_{+} +\frac {3 R^{(p)}_{0}}{2\sqrt{5}}\right) \frac {\partial}{\partial r} \right] R_B\Bigg|^2
\end{equation}

\begin{figure}[t]
\centering
\includegraphics[width=.49\textwidth]{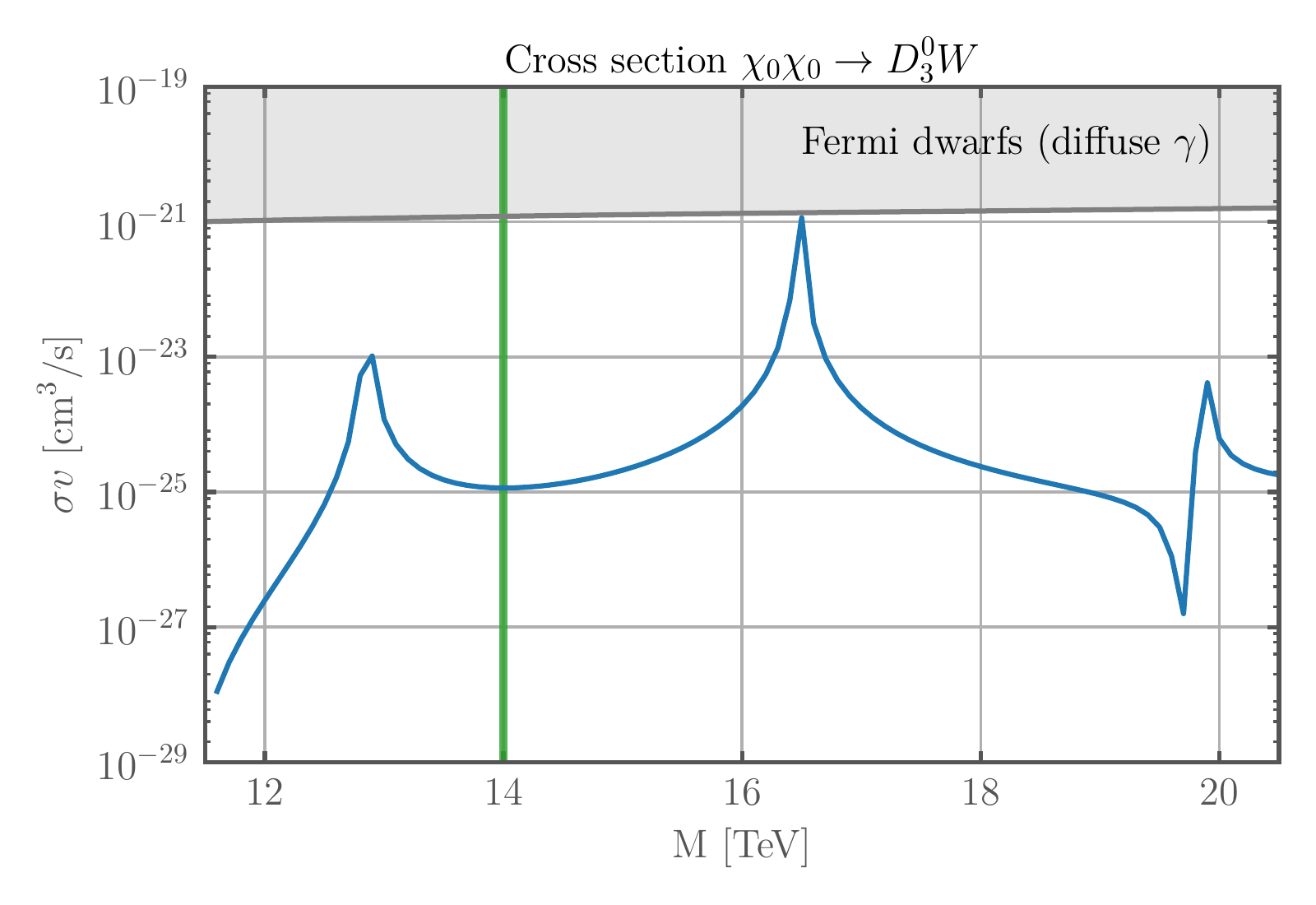}~
\includegraphics[width=.49\textwidth]{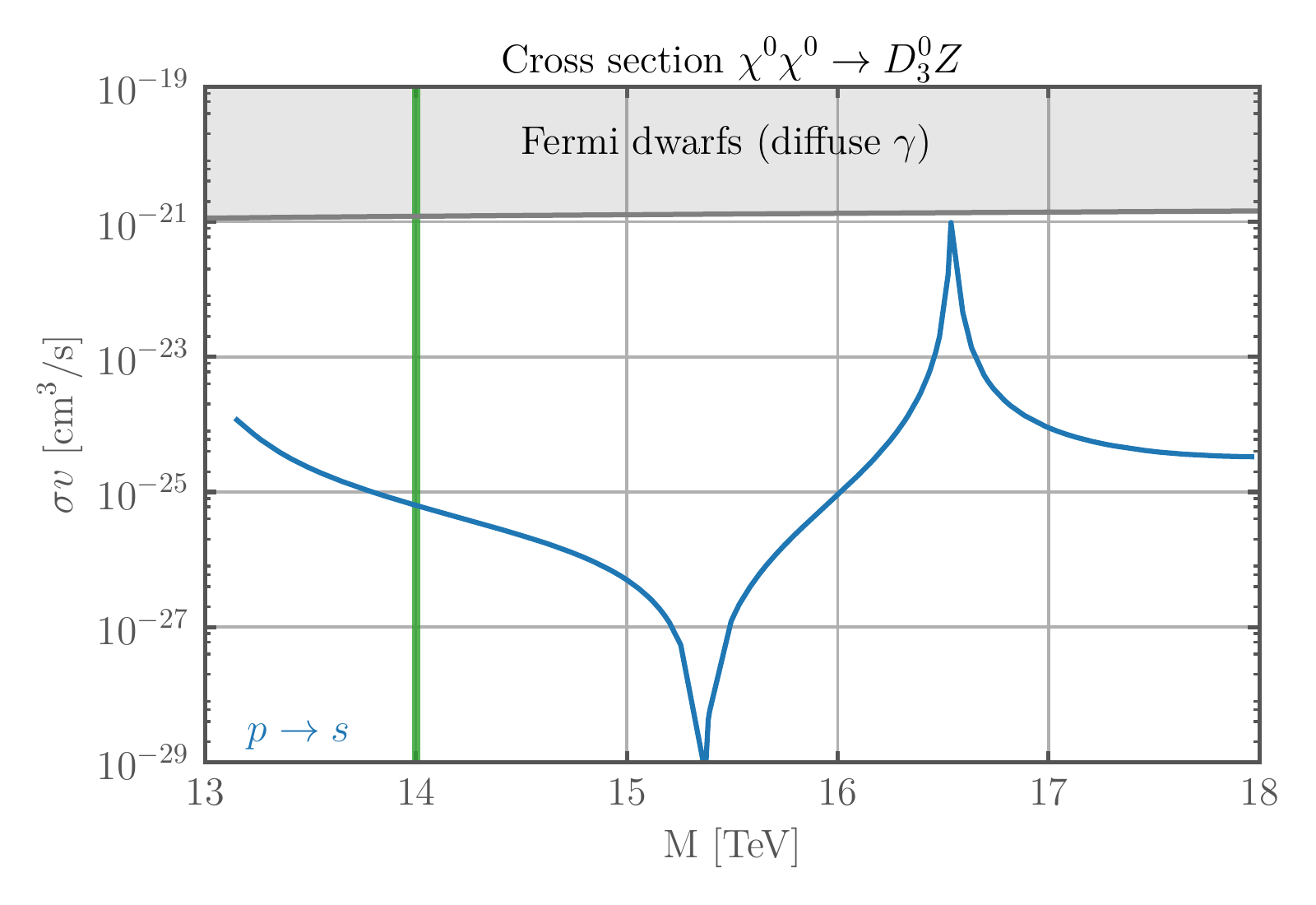}
\caption{\label{fig:BS5pletW}  Cross section for bound state formation
  of s-wave  isospin-triplet (1$s_3$) bound state by emission of $W$ (left) or $Z$ (right) bosons.}
\label{fig:DF}
\end{figure}

The cross sections are reported in Fig.(\ref{fig:BS5pletW}). For the nominal value of the thermal mass the bounds do not give significant constraints.

 
\section{Conclusions}
\label{sec:conclusions}

In this work we studied the formation of  DM bound states and their potential impact on astrophysical signals of DM.  
Bound state formation is expected to be generic in theories where DM has strong interactions,  for instance when DM is the lightest baryon of a confining dark QCD-like gauge theory.
Our results are complementary and generalize in the strongly coupled regime Ref. \cite{Mitridate:2017izz} where production of perturbative 
hydrogen-like bound states was studied in weakly-coupled non-abelian gauge theories. 

The physics of strongly-coupled DM bound states is analogous to that of deuterium in the SM, where the smallness of the binding energy
allows us to compute the relevant cross sections using the effective range expansion, without a detailed knowledge of the underlying nuclear potential.
If DM is charged under the SM electroweak group, bound states of DM  can form 
through the emission of a monochromatic photon, or electroweak gauge bosons if kinematically allowed.

At the technical level we showed that the presence of a short-range (nuclear) component to the
potential often invalidates the ubiquitous  factorization of the cross section into a Sommerfeld factor, which encodes the long-range effect due to
the electroweak potential, times a short-distance cross section. This failure obviously arises due to the finite extent of the bound states, but more
importantly is due to the modification of the spectrum of zero energy bound states by the short distance potential. 
The latter effect would also impact DM annihilation if the DM has a strong short-distance potential.
In light of this the relevant cross section for bound-state DM should be safely computed using explicit wavefunctions 
in order to reliably capture the interplay between the `short-distance' and `long-distance' physics.

After providing general formulae for bound-state formation, we considered in detail the minimal model where DM is the neutral
component of an SU(2)$_L$-triplet baryon. Formation of dark deuterium
gives rise to monochromatic and diffuse photon signals can be constrained by existing FERMI and HESS measurements, 
and will be further tested by future experiments such as HERD. 

With the tools developed in this work we also revisited SU(2)$_L$-quintuplet Minimal DM, updating previous studies.
Besides its direct annihilation to SM particles, bound state formation leads to novel signatures for this DM candidate 
that are unavoidable and complementary although currently consistent existing measurements for M=14 TeV.

Our formalism, which allows us to compute the formation of shallow bound states on the emission of light quanta in the
strongly-coupled regime, is easily extended to other scenarios. For example  if Yukawa interactions with the SM Higgs exist, bound-state formation can
proceed through Higgs emission.  Emission of light particles coupled to the SM at sufficiently low velocities could also be studied in a similar way. 
We leave these and other questions to future work.

{\small
\subsubsection*{Acknowledgements}
We thank Filippo Sala for useful discussions. 
This work is supported by MIUR grants PRIN 2017FMJFMW and 2017L5W2PT, Ente Cassa di Risparmio di Firenze and INFN grant STRONG.}

\appendix

\section{Factorized cross section}\label{app:Factorized}

In this appendix we provide explicit formulae for formation of shallow bound states that are applicable 
when the long distance effects can be factorized in the Sommerfeld enhancement factor. 
In this case the cross section can be cast in form similar to the annihilation of DM. By solving the Schroedinger equation 
for the initial state without strong interactions we can define the matrix,
\begin{equation}
A_{ia}= \frac {\psi_i(0)}{\psi_a^0(0)}
\end{equation}
where $\psi_a^0$ is the initial state free wavefunction in the channel $a$. The matrix $A_{ia}$ encodes the Sommerfeld enhancement
and is identical to one computed for annihilation.

The full cross section for the process $a\to i \to f$ reads \cite{Hisano:2004ds}
\begin{equation}
\sigma_{a} = c_a (A^\dag \cdot \Gamma^f \cdot A)_{aa} 
\label{eq:hisano}
\end{equation}
where $\Gamma^f$ is a generalized the cross section for the nuclear process and $c_a=2 (1)$ for Majorana (Dirac) particles.
Note that  diagonal components of $\Gamma^f$ describe physical cross sections while the off-diagonal components corresponds to interference effects.

Since the initial states are charge eigenstates it is useful to give the annihilation matrices in that basis. 
For magnetic transition the generalized cross section matrix appearing in (\ref{eq:hisano}) can be written as \cite{Cirelli:2007xd},
\begin{equation}
\Gamma_{ij,i'j'}^{\rm mag}= N_{ii'}N_{jj'} \kappa^2\frac {2^8}{g_N^2}\sigma_0    \left(1- \frac{M_a^2}{E_B^2}\right)^{\frac 3 2}\left(\frac {E_B} {M}\right)^{\frac 3 2}
\sum_{\mathbf{r}}(1-a_{\mathbf{r}}\gamma_{\mathbf{r}'}){\rm CG}_{\mathbf{r}, ii'}^M  C_{\cal J}^{a M M'}
\sum_{\mathbf{r}}(1-a_{\mathbf{r}}\gamma_{\mathbf{r}'}){\rm CG}_{\mathbf{r}, jj'}^M  C_{\cal J}^{a M M'}
\end{equation}
where $N_{ij}=1$ for $i\ne j$ and $N_{ij}=1/\sqrt{2}$ for $i=j$.
The group theory factors take into account that charge eigenstates are not mass eigenstates from the point of view of strong interactions.

For electric cross sections instead,
\begin{equation}
\Gamma_{ij,i'j'}^{\rm el}=v_{\rm rel}^2 N_{ii'}N_{jj'} \frac {2S+1}3 \frac {2^6}{g_N^2}\sigma_0    \sqrt{1- \frac{M_a^2}{E_B^2}} \left(1+\frac {M_a^2} {2E_{B_{\mathbf{3}}}^2}\right)\sqrt{\frac {M}{E_B}}
\sum_{\mathbf{r}}{\rm CG}_{\mathbf{r}, ii'}^M  C_{\cal J}^{a M M'}
\sum_{\mathbf{r}}{\rm CG}_{\mathbf{r}, jj'}^M  C_{\cal J}^{a M M'}
\end{equation}

\subsection{Weak triplet}

As an example we here report the generalized cross sections matrices for the production of the s-wave nuclear bound state $D_3$ 
from an initial state with two neutral nucleons $V^0$ in the triplet representation of $SU(2)_L$.
For the magnetic transition the initial state is s-wave with spin-0. One finds,
\begin{eqnarray}
&\Gamma_{D^0_{\mathbf{3}}\gamma}^{\rm mag} = \kappa^2 \frac {2^7}{9}\frac{\pi \alpha_{\rm em}}{M^2}\left( \frac {E_{B_{\mathbf{3}}}}{M}\right)^{\frac 3 2}\nonumber \\
&\times \begin{pmatrix}
\left(\frac 3 2- a_{\mathbf{1}} \gamma_\mathbf{3}-\frac 1 2 a_{\mathbf{5}} \gamma_\mathbf{3}\right)^2 & \left(\frac 3 2- a_{\mathbf{1}} \gamma_\mathbf{3}- \frac 1 2 a_{\mathbf{5}} \gamma_\mathbf{3}\right)\left(a_{\mathbf{5}}\gamma_{\mathbf{3}} - a_{\mathbf{1}} \gamma_{\mathbf{3}}\right)/\sqrt{2}\\
\left(\frac 3 2- a_{\mathbf{1}} \gamma_\mathbf{3}- \frac 1 2 a_{\mathbf{5}} \gamma_\mathbf{3}\right)\left(a_{\mathbf{5}}\gamma_{\mathbf{3}} - a_{\mathbf{1}} \gamma_{\mathbf{3}}\right)/\sqrt{2} & \left(a_{\mathbf{5}}\gamma_{\mathbf{3}} - a_{\mathbf{1}} \gamma_{\mathbf{3}}\right)^2/2 \\
\end{pmatrix}\nonumber \\
&\Gamma_{D^0_{\mathbf{3}}Z}^{\rm mag} = \kappa^2 \frac {2^7}{9}\frac{\pi \alpha_{2}c_W^2}{M^2}\left( \frac {E_{B_{\mathbf{3}}}}{M}\right)^{\frac 3 2}\left(1- \frac {M_Z^2}{E_{B_{\mathbf{3}}}^2}\right)^{\frac 3 2}\nonumber \\
&\times \begin{pmatrix}
\left(\frac 3 2- a_{\mathbf{1}} \gamma_\mathbf{3}-\frac 1 2 a_{\mathbf{5}} \gamma_\mathbf{3}\right)^2 & \left(\frac 3 2- a_{\mathbf{1}} \gamma_\mathbf{3}- \frac 1 2 a_{\mathbf{5}} \gamma_\mathbf{3}\right)\left(a_{\mathbf{5}}\gamma_{\mathbf{3}} - a_{\mathbf{1}} \gamma_{\mathbf{3}}\right)/\sqrt{2}\\
\left(\frac 3 2- a_{\mathbf{1}} \gamma_\mathbf{3}- \frac 1 2 a_{\mathbf{5}} \gamma_\mathbf{3}\right)\left(a_{\mathbf{5}}\gamma_{\mathbf{3}} - a_{\mathbf{1}} \gamma_{\mathbf{3}}\right)/\sqrt{2} & \left(a_{\mathbf{5}}\gamma_{\mathbf{3}} - a_{\mathbf{1}} \gamma_{\mathbf{3}}\right)^2/2 \\
\end{pmatrix}\nonumber \\
&\Gamma_{D^+_{\mathbf{3}}W}^{\rm mag} = \kappa^2 \frac {2^6}{9}\frac{\pi \alpha_2}{M^2}\left( \frac {E_{B_{\mathbf{3}}}}{M}\right)^{\frac 3 2}\left(1- \frac {M_W^2}{E_{B_{\mathbf{3}}}^2}\right)^{\frac 3 2}\nonumber \\
&\times \begin{pmatrix}
\left(\frac 3 2- 2a_{\mathbf{1}} \gamma_\mathbf{3}-\frac 1 2 a_{\mathbf{5}} \gamma_\mathbf{3}\right)^2 & \left(\frac 3 2- 2a_{\mathbf{1}} \gamma_\mathbf{3}- \frac 1 2 a_{\mathbf{5}} \gamma_\mathbf{3}\right)\left(3-2 a_{\mathbf{1}}\gamma_{\mathbf{3}} - a_{\mathbf{5}} \gamma_{\mathbf{3}}\right)/\sqrt{2}\\
\left(\frac 3 2- 2 a_{\mathbf{1}} \gamma_\mathbf{3}- \frac 1 2 a_{\mathbf{5}} \gamma_\mathbf{3}\right)\left(3-2a_{\mathbf{1}}\gamma_{\mathbf{3}} - a_{\mathbf{5}} \gamma_{\mathbf{3}}\right)/\sqrt{2} & \left(3-2 a_{\mathbf{1}}\gamma_{\mathbf{3}} - a_{\mathbf{5}} \gamma_{\mathbf{3}}\right)^2/2 \\ 
\end{pmatrix}\nonumber \\
\end{eqnarray}

For electric transitions the initial state is p-wave with spin-1. One finds,
\begin{eqnarray}
\begin{split}
&\Gamma_{D^0_{\mathbf{3}}\gamma}^{\rm el} = \frac {v_{\rm rel}^2}4 \frac {2^7}{9}\frac{\pi \alpha_{\rm em}}{M^2}\sqrt{\frac {M}{E_{B_{\mathbf{3}}}}}
\times \begin{pmatrix}
\frac 9 4 & 0 \\
0 &0  \\
\end{pmatrix} \\
&\Gamma_{D^+_{\mathbf{3}}W}^{\rm el} = \frac {v_{\rm rel}^2}4 \frac {2^6}{9}\frac{\pi \alpha_2}{M^2}\sqrt{\frac {M} {E_{B_{\mathbf{3}}}}}\sqrt{1- \frac{M_W^2}{E_{B_{\mathbf{3}}}^2}}  \left(1+\frac{M_W^2}{2E_{B_{\mathbf{3}}}^2}\right)
\times \begin{pmatrix}
\frac 9 4 & \frac 9 {2\sqrt{2}} \\
\frac 9 {2\sqrt{2}} & \frac 9 2  \\
\end{pmatrix}\\
&\Gamma_{D^0_{\mathbf{3}}Z}^{\rm el} = \frac {v_{\rm rel}^2}4 \frac {2^7}{9}\frac{\pi \alpha_2 c_W^2}{M^2}\sqrt{\frac {M} {E_{B_{\mathbf{3}}}}}\sqrt{1- \frac{M_Z^2}{E_{B_{\mathbf{3}}}^2}}  \left(1+\frac{M_Z^2}{2E_{B_{\mathbf{3}}}^2}\right)
\times \begin{pmatrix}
\frac 9 4 &0 \\
0 & 0  \\
\end{pmatrix}
\end{split}
\end{eqnarray}

\section{Sommerfeld enhancement and phase shifts}\label{appendixB}
In this appendix we review the computation of Sommerfeld factors. 
Since we only consider rotationally invariant interactions, the angular and radial part of the wavefunction separate from each other. The scattering process, at large distances, is characterised by the asymptotic expression
\be\label{asymptotics}
\psi_{ia}^{\rm scattering}\stackrel{r\to \infty}{=} \delta_{ia} e^{i p_a z} + f_{ia}(\theta) \frac{e^{i p_i r}}{r}\,,
\ee 
where we fix the axis of the incoming plane wave to be $\vec{p}\cdot \vec{r}= p r \cos\theta= p z$.
In general the plane wave along $z$ can be decomposed as $e^{ikz}=\sum i^\ell Y_{\ell,0} (\theta)\sqrt{4\pi (2\ell+1)}j_\ell(kr)$, where $j_\ell$ are the spherical Bessel functions and $Y_{\ell,m=0}(\theta)=\sqrt{(2\ell+1)/4\pi}P_\ell(\cos\theta)$ the spherical harmonics. Therefore, by exploiting the behavior $j_\ell(\rho)\to \sin(\rho-\ell\pi/2)/\rho$ for large $\rho$,  we get
\be\label{asymptotic}
\psi_{ia}^{\rm scattering}\stackrel{r\to \infty}{=} \frac{e^{-i p_a r}}{2i p_a r} \sum_\ell (-1)^{\ell+1}\delta_{ia} (2\ell+1)P_\ell(\cos\theta) +  \frac{e^{i p_i r}}{2i p_a r} \bigg[ \sum_\ell \delta_{ia} (2\ell+1)P_\ell(\cos\theta) + 2i p_a f_{ia}(\theta)\bigg]\,.
\ee
The label $a$ represents the initial wave packet, incoming with momentum $p_a$, while $i$ labels the possible final states, and in general we allow for $i\neq a$. Exploiting rotational symmetry we can write down the full solution as
\be\label{scattering-state}
\psi_{ia} =  \sum_{\ell} Y_{\ell,0}(\theta) R_\ell^{ia}(r) \equiv  \sum_{\ell} Y_{\ell,0}(\theta) \frac{u_\ell^{ia}(r)}{r} = \sum_{\ell} \sqrt{\frac{2\ell+1}{4\pi}} P_\ell(\cos\theta) \frac{u_\ell^{ia}(r)}{r}\,,
\ee
where we have used the fact that $m=0$. Then we need to find the correct asymptotic behavior that matches eq.~\eqref{asymptotic}. The asymptotic behaviour of the reduced radial functions that matches eq.\eqref{asymptotics} is given by
\be\label{chi-asymp}
 u_\ell^{ia}\to \frac{\sqrt{4\pi(2\ell+1)}}{2ip_a} \bigg[ (-1)^{\ell+1} \delta_{ia} e^{-i p_a r} + S_\ell^{ia} e^{i p_i r}\bigg]\,.
\ee
The boundary conditions at the origin are such that the regular solutions are chosen $u_\ell^{ia}\propto r^{\ell+1}$.
Notice that the asymptotic matching \eqref{chi-asymp} can be realised by requiring that at (numerical) infinity $r_\infty$
\be
\frac{du^\ell_{ia}}{dr}(r_\infty)- i p_i u^\ell_{ia}(r_\infty)= (-1)^\ell \sqrt{4\pi(2\ell+1)} \delta_{ia} e^{-i p_i r_{\infty}}\,,
\ee
often for the off-shell mode $i\neq a$, the condition $u_{ia}(r_\infty)=0$ is more stable. These two conditions allows us to compute the full wavefunction for a given initial state potential. We can then extract other information, such as Sommerfeld factors and phase shifts.

\paragraph{Sommerfeld factors for any $\ell$.}
The Sommerfeld factor for generic $\ell$ can be extracted from appropriate number of derivatives of the reduced wavefunction (recall that the Sommerfeld factor is defined as the ratio of $\ell$-wave part of the wavefunction at the origin and the unperturbed wave along $z$):
\be\label{eq:sommerfeld-general}
A_{ia}\big|_{\ell}\equiv \frac{\psi_{ia}(0)}{\big[e^{ikz}\big]_\ell (0)}= \frac{(2\ell +1)!!}{i^{\ell}\sqrt{4\pi(2\ell+1)} (\ell +1)! } \frac{1}{p_a^\ell}\, \lim_{r\to 0}\, \frac{d^{\ell+1}}{ d r^{\ell+1}}u_{ia}(r)\,.
\ee
In the above expression we have made use of the fact that spherical Bessel functions expand as $j_\ell(\rho)\to \rho^\ell/(2\ell+1)!!$ for small $\rho$, and exploited the knowledge that $u\sim r^{\ell +1}$ at the origin.
For the specific case of $s$ and $p$-wave scattering, we get
\be
A_{ia}\big|_{s}=\frac{1}{\sqrt{4\pi}} \frac{du_{ia}}{dr}\big|_{r=0}\,,\quad  A_{ia}\big|_{p}=\frac{-i}{\sqrt{4\pi}}\frac{\sqrt{3}}{2p_a}  \frac{ d^2u_{ia}}{dr^2}\big|_{r=0} \,.
\ee
\paragraph{Phase shifts for any $\ell$.}
The phase shift of the incoming wave packet can be computed from the amplitude of the out-going function.
From eq.~\eqref{asymptotics} it follows that,
\begin{equation}
\frac{du^\ell_{ia}}{dr}(r_\infty)+ i p_i u^\ell_{ia}(r_\infty)= (-1)^\ell \sqrt{4\pi(2\ell+1)} S_{ia} e^{i p_i r_{\infty}}\,,
\end{equation}
which in turn allows us to compute $f(\theta)$ as
\be
f_{ia}(\theta)=\frac{2\ell+1}{2ip_a}P_\ell(\cos\theta) (S_{ia}-1) = \frac{(2\ell+1)P_\ell(\cos\theta)}{p_a} e^{i\delta_{ia}} \sin(\delta_{ia})\,\quad  \mathrm{for\,}\, i=a\,.
\ee
The cross section for elastic scattering then given by,
\begin{equation}
\sigma_{\rm el}=\frac{\pi (2\ell+1)}{p_a^2}| S_{ia}-1|^2=\frac{4\pi (2\ell+1)}{p_a^2}| e^{i\delta_{ia}} \sin(\delta_{ia})|^2\,.
\end{equation}

\section{Modelling the nuclear potential}\label{app:nucl}
The computation of bound state formation with wavefunctions requires the choice of a potential.
In regime of shallow bound states the result is only weakly sensitive to details of the potential so one 
can choose the most convenient.

As from textbooks on  quantum mechanics \cite{Schiff}, a convenient parametrization  is the spherical well potential, with tunable parameters. Since the strong interactions are isospin symmetric, in each channel of spin and weak isospin we have a corresponding potential $V(r)=-V_N \theta(r_0 - r)$, with $V_N>0$ and where $r_0$ is the range of the interaction, that we assume to be related to the dark pion mass by $r_0 \sim M_\pi^{-1}$. The only difference among different channels will be the depth of the well. 
This is the potential that is included in the numerical simulation. By knowing the spectrum of dark bound states in a given channel of spin and isospin, is then simple to include this effects in the initial state wavefunctions, by simply choosing just $V_N$ to reproduce a given binding energy. For a fixed range $r_0$ is always possible to tune the depth of the well $V_N$ in such a way to have only one bound state with zero angular momentum and arbitrarily small binding energy. The normalized reduced wavefunction with binding energy $E_B=\gamma^2/M$ is explicitly given by
\be
u_B(r) =\frac{\sqrt{2\gamma}}{{\sqrt{1+r_0 \gamma}}}  \bigg[  \sin(\kappa r)\theta(r_0-r) + \sin(\kappa r_0)e^{-\gamma(r-r_0) }\theta(r-r_0)\bigg]\,, \quad \kappa\equiv\sqrt{MV_N -\gamma^2}\,,
\ee
while the binding energy is given implicitly by the solution of
\be\label{bindingenergy}
\kappa \cot(\kappa r_0)=-\gamma\,.
\ee
When the potential supports a single shallow bound state the equation above can be solved as,
\begin{equation}
\frac {V_N}{M}=\frac {\pi^2}{4} x_0^2+ 2 \sqrt{\frac {E_B} M} x_0+\left(1 -\frac 4 {\pi^2}\right) \frac {E_B} M+  \dots
\label{Bapprox}
\end{equation}
where $x_0=1/(M r_0)$.
Some benchmarks values are,
\begin{center}
\begin{tabular}{c||c|c}
$E_B/M$ &  $V_N/M (x_0=0.3)$ & $V_N/M ( x_0=1)$\\ 
\hline 
0.01 &  0.29 & 2.67\\
0.05 &  0.39 & 2.94 \\
0.1 &  0.48 & 3.16 \\
\end{tabular}
\end{center}
In our numerical simulation we fix $x_0=1$ corresponding to large pion masses $M_\pi\sim M$ and $E_B/M$ in a range between $0.1$ and $0.01$. 

\paragraph{$s$-wave process.~\\}
For s-wave the normalized positive energy wavefunction is
\be
u_s(r)=\frac{\sqrt{4\pi}}{p} \bigg[\frac{\sin(K r)}{ \sin(K r_0)} \sin(p r_0+\delta_0) \theta(r_0-r)+ \sin(p r+\delta_0) \theta(r-r_0)\bigg]\,,\quad K\equiv\sqrt{MV_N + p^2},
\label{bs-well}
\ee
The phase shift is determined by regularity of the wavefunction at $r=r_0$,
\be
\tan(p r_0 +\delta_0)=\frac{p}{K}\tan{K r_0}  \longrightarrow  \delta_0= p \left(\frac{\tan (\sqrt{MV_N} r_0)}{\sqrt{M V_N}} -r_0\right )+ O(p^2)
\ee
In the limit of small binding energy using \eqref{bindingenergy} one finds
\be
\delta_0= -\frac{p}{\gamma} - p r_0 + O(p^2)
\label{delta0}
\ee
This result can be directly derived using the effective range expansion. The amplitude for elastic scattering has the general form
\begin{equation}
{\cal A}= \frac {4\pi} M \frac 1 {p \cot \delta_{\ell} -i p}
\end{equation}
In the  low velocity regime the scattering phase admit the expansion $p^{2\ell+1} \cot \delta_{\ell}= - a_l^{-2 \ell-1}+{\cal O}(p^2)$ where $a_\ell$ is the scattering length.
Thus the amplitude has a pole for $i \sqrt{M E_B}=1/a$ that can be trusted for $E_B \ll M$ and eq. (\ref{delta0}) follows.

From the wavefunction above we can extract the Sommerfeld factor using eq.~\eqref{eq:sommerfeld-general}. The result is
\begin{equation}
A_s=\frac K p \frac {\sin(p r_0+\delta_0)}{\sin (K
  r_0)}   \longrightarrow {\rm SE}_0=|A_s|^2\approx \frac { V_N M a_0^2}{1+p^2 a_0^2}
\end{equation}
where we have used $p\cot\delta_0\approx -1/a_0$ effectively resumming $a_0 p$ to all orders. 
 For vanishing binding energy the SE diverges in the low velocity regime giving rise to the peaks in the cross section.

For magnetic transition the overlap integral between initial and final state is given by,
\be\label{eq-bethe}
\int_0^\infty dr u_s u_B = \frac {\sqrt{8\pi\gamma_f}(p \cos \delta_0+\gamma_f \sin \delta_0)}{p(p^2+\gamma_f^2)}+O(r_0)
\approx -\frac{\sqrt{8\pi}  (1-a_i \gamma_f)}{\gamma_f^{3/2}} + O(r_0)\,.
\ee
where in the last step we expanded in the low velocity limit. Substituting in eq. (\ref{eq:magQM}) this gives the magnetic cross section in eq. (\ref{xsec:magnetic}).
Note that to leading order the result is independent of $r_0$ so that the computation can be simply performed in the limit $r_0=0$  \cite{Bethe:1949yr}.

\paragraph{$p$-wave process.}~\\
Let us repeat the exercise for $\ell=1$. The condition for a zero energy bound state is now
\begin{equation}
V_N = \frac {\pi^2}{M r_0^2}\,.
\end{equation}
The explicit solution is
\be\label{p-well}\small
u_p(r)= N\bigg( \frac {\sin (K r)}{K r} -\cos (K r)\bigg) \theta(r_0 - r) +  \frac {\sqrt{12 \pi}}p \bigg[ \cos \delta_1 \bigg( \frac {\sin (p r)}{p r} -\cos (p r)\bigg)- \sin \delta_1 \bigg( \frac {\cos (p r)}{p r} +\sin (p r)\bigg)\bigg] \theta(r-r_0)
\ee
where $N$ and $\delta_1$ are determined by matching the wavefunctions and the first derivative at the boundary. 
The SE amplitude is then
\begin{equation}
A_p=-  i \frac {N K^2}{\sqrt {12 \pi} p} \,.
\end{equation}
For shallow bound states we can get an explicit formula as follows. To leading order $\tan \delta_1= - (p a_1)^3$. 
By looking at the elastic amplitude $a\approx 1/\sqrt{M E_B}$. Plugging in the formula above we find
\begin{equation}
{\rm SE}_1=|A_p|^2\approx \frac{a_1^6 M^3 V_N^3}{\pi^{2} (1+ a_1^6 p^6)}\,.
\end{equation}
Note the very different energy dependence compared to s-wave processes.

We can compute the cross section for the production of an s-wave shallow bound state from a p-wave initial state
using the explicit wavefunctions (\ref{p-well}) and (\ref{bs-well}). The relevant matrix element is given by
\begin{equation}
\int_0^\infty dr\, r\, u_p \partial_r \left(\frac {u_B}r\right)=\,2p  \frac {\sqrt{6\pi\gamma_f}}{p^2 +\gamma_f^2}+ O(r_0)
\end{equation}
that does not depend on the initial scattering length to leading order.

%

\section{Modified Variable Phase
  Method}\label{appendixD}\pagestyle{plain}
(Based on procedure in \cite{Ershov:2011zz} as modified by \cite{Beneke:2014gja}, with modified boundary
condition from \cite{Asadi:2016ybp}.)

We want to solve the radial Schrodinger equation for the reduced
wavefunction $u_\ell$,
\begin{equation}\label{eq:RadialSchrodinger}
[u_\ell''(x)]_{in}+
\left(1+\frac{\ell(\ell+1)}{x^2}\right)[u_\ell(x)]_{in}=\frac{1}{M\beta^2}\sum_{j=1}^nV_{ij}(x)
[u_\ell(x)]_{jn}
\end{equation}
where $x$ is a dimensionless radial variable defined as $x=M\beta
r$, and $i,n=1,\cdots,N$ corresponding to the $N$ different components
of the wavefunction.  For the EW triplet case, $N=2$ for the
charged-charged and neutral-neutral components of the (spin-0,
charge-0) wavefunction.  

We split up the potential into an asymptotic part and a
short-distance part as $V(x)=V^\infty+\hat{V}(x)$, where 
\[
V^\infty=\lim_{x\to \infty} V(x)
\]
We will use the regular and irregular solutions ($f_i(x)$ and
$g_i(x)$, respectively) of the `free'
Schr\"odinger equation:
\begin{equation}
\left[\frac{d^2}{dx^2} + 1 -\frac{V^\infty_{ii}}{M\beta^2}- \frac{\ell(\ell+1)}{x^2}\right]\left[\begin{array}{c}f_i(x)\\g_i(x)\end{array}\right]=0
\end{equation}
The free solutions are normalized such that the
Wronskian, 
\begin{equation}
f_i(x)g'_i(x)-f'_i(x)g_i(x)=-1\,.
\end{equation}
i.e.
\[
f_i(x)=\sqrt{\frac{\pi x}{2}}J_{\ell+\frac{1}{2}}(\hat{k}_i x)
\qquad g_i(x)=-\sqrt{\frac{\pi
    x}{2}}\left[Y_{\ell+\frac{1}{2}}(\hat{k}_i x) - i
  J_{\ell+\frac{1}{2}}(\hat{k}_i x)\right]
\]
for dimensionless wavenumber
$\hat{k}_i=(1-V^\infty_{ii}/(M\beta^2))^{1/2}$.

The variable phase ansatz states that we can write the regular solution to the full Sch\"odinger equation as linear
combinations of the `free' solutions for unknown functions $\alpha_{in}(x),\,\beta_{in}(x)$
\begin{equation}
[u_\ell(x)]_{in} = f_i(x)\alpha_{in}(x)- g_i(x)\beta_{in}(x)
\end{equation}
 for each $\ell$, with boundary
conditions $\alpha_{in}(0)=\delta_{in}$ and $\beta_{in}(0)=0$.  We
have doubled the degrees of freedom, so we also impose the following
constraint 
\begin{equation}
f_i(x)\alpha'_{in}(x)=g_i(x)\beta'_{in}(x)
\end{equation}
which reduces the $N$ second-order equations,
Eq.~\eqref{eq:RadialSchrodinger} to a system of $2N$ first-order coupled
ODEs:
\begin{eqnarray}\label{eq:2NLinear}
\alpha'_{in}(x)&=&\frac{g_i(x)}{M\beta^2}\sum_{j=1}^N
  \hat{V}_{ij}(x)[u_\ell(x)]_{jn}\nonumber\\
\beta'_{in}(r)&=&\frac{f_i(x)}{M\beta^2}\sum_{j=1}^N
  \hat{V}_{ij}(x)[u_\ell(x)]_{jn}
\end{eqnarray}
and $\beta(x)$ is linearly dependent on $\alpha(x)$: 
\begin{equation}\label{eq:BetaDependent}
\beta_{in}(x)=\sum_{j=1}^{N}\mathcal{O}_{ij}(x)\alpha_{jn}(x)
\end{equation}
making
\[
[u_l(x)]_{in}=f_i(x)\alpha_{in}(x)-g_i(x) \sum_{j=1}^{N}\mathcal{O}_{ij}(x)\alpha_{jn}(x)\,.
\]

Using \eqref{eq:BetaDependent} in
\eqref{eq:2NLinear} we obtain:
\begin{equation}\label{eq:OPrime}
\mathcal{O}'_{ij}=\frac{1}{M\beta^2}\sum_{k,m=1}^N\left(\delta_{ik}f_k-\mathcal{O}_{ik}g_k\right) \hat{V}_{km}\left(f_m\delta_{mj}-g_m\mathcal{O}_{mj}\right)
\end{equation}
where the argument $x$ has been suppressed everywhere for brevity.

Introducing matrix $N_{ij}$ minimizes
numerical convergence issues (see \cite{Ershov:2011zz} for more details):
\[
N_{ij}=f_ig_i\delta_{ij}-g_i\mathcal{O}_{ij}g_j
\]
Differentiating the above, and substituting for $\mathcal{O}'$ from
\eqref{eq:OPrime}, we can derive that $N_{ij}$ satisfies the following
differential equation:
\begin{equation}\label{eq:NPrime}
N'_{ij}=\delta_{ij}+\left(\frac{g'_i}{g_i}+\frac{g'_j}{g_j}\right)N_{ij}- \frac{1}{M\beta^2}\sum_{k,m=1}^NN_{ik}\hat{V}_{km}N_{mj}
\end{equation}
Following \cite{Ershov:2011zz} we can write
\[
[u_\ell]_{in}=\sum_{j=1}^NN_{ij}\frac{\hat{\alpha}_{jn}}{g_n}
\]
where $\hat{\alpha}_{in}=\frac{g_n}{g_i}\alpha_{in}$.  Differentiating this, we
obtain a differential equation for $\hat{\alpha}_{in}$ that is
logarithmically stable:
\begin{equation}\label{eq:AlphaTildePrime}
  \hat{\alpha}'_{in}=\left(\frac{g'_n}{g_n}-\frac{g'_i}{g_i}\right)\hat{\alpha}_{in}+\frac{1}{M\beta^2}\sum_{j,k=1}^N\hat{V}_{ik}N_{kj}\frac{\hat{\alpha}_{jn}}{g_n} 
\end{equation}
Imposing physical boundary conditions for the regular solution as $x\to 0$
\[
\lim_{x\to
  0}[u_\ell(x)]_{in}=\frac{1}{2\ell+1}x^{\ell+1}\delta_{in}\qquad\lim_{x\to 0}[u'_\ell(x)]_{in}=\frac{\ell+1}{2\ell+1}x^l\delta_{in}
\]
translate into
\[
\lim_{x\to
  0}\alpha^{(\ell)}_{in}(x)=\frac{(2\ell-1)!!}{\hat{k}_i^{\ell+1/2}}\delta_{in}\qquad\qquad
\lim_{x\to 0}\beta^{(\ell)}_{in}(x)=0
\]
which, in turn implies
\begin{equation}\label{eq:BoundaryConditions}
\lim_{x\to
  0}N_{ij}(x)=\frac{x}{2\ell+1}\delta_{ij}\qquad\textrm{and}\qquad
\lim_{x\to 0}\hat{\alpha}_{in}(x)=(2\ell-1)!!\,\delta_{in}
\end{equation}
So to solve for the physical wavefunctions, we need to solve
first-order differential equations \eqref{eq:NPrime} and
\eqref{eq:AlphaTildePrime}, with boundary conditions \eqref{eq:BoundaryConditions}.

Unfortunately imposing $\hat{\alpha}$ boundary conditions at small
$x$ leads to numerical instabilities, and so we impose instead
\[
\lim_{x\to\infty}\hat{\alpha}_{in}(x)=\delta_{in}\,.
\]
\pagestyle{plain}
\bibliographystyle{jhep}
\small
\bibliography{biblio}

\providecommand{\href}[2]{#2}\begingroup\raggedright\begin{thebibliography}{10}

\bibitem{CyrRacine:2012fz}
F.-Y. Cyr-Racine and K.~Sigurdson, {\it {Cosmology of atomic dark matter}},
  {\em Phys. Rev.} {\bf D87} (2013), no.~10 103515,
  [\href{http://arxiv.org/abs/1209.5752}{{\tt arXiv:1209.5752}}].

\bibitem{Mitridate:2017izz}
A.~Mitridate, M.~Redi, J.~Smirnov, and A.~Strumia, {\it {Cosmological
  Implications of Dark Matter Bound States}},  {\em JCAP} {\bf 1705} (2017),
  no.~05 006, [\href{http://arxiv.org/abs/1702.01141}{{\tt arXiv:1702.01141}}].

\bibitem{Harz:2018csl}
J.~Harz and K.~Petraki, {\it {Radiative bound-state formation in unbroken
  perturbative non-Abelian theories and implications for dark matter}},  {\em
  JHEP} {\bf 07} (2018) 096, [\href{http://arxiv.org/abs/1805.01200}{{\tt
  arXiv:1805.01200}}].

\bibitem{Cirelli:2007xd}
M.~Cirelli, A.~Strumia, and M.~Tamburini, {\it {Cosmology and Astrophysics of
  Minimal Dark Matter}},  {\em Nucl. Phys.} {\bf B787} (2007) 152--175,
  [\href{http://arxiv.org/abs/0706.4071}{{\tt arXiv:0706.4071}}].

\bibitem{Redi:2018muu}
M.~Redi and A.~Tesi, {\it {Cosmological Production of Dark Nuclei}},  {\em
  JHEP} {\bf 04} (2019) 108, [\href{http://arxiv.org/abs/1812.08784}{{\tt
  arXiv:1812.08784}}].

\bibitem{Antipin:2015xia}
O.~Antipin, M.~Redi, A.~Strumia, and E.~Vigiani, {\it {Accidental Composite
  Dark Matter}},  {\em JHEP} {\bf 07} (2015) 039,
  [\href{http://arxiv.org/abs/1503.08749}{{\tt arXiv:1503.08749}}].

\bibitem{Kribs:2016cew}
G.~D. Kribs and E.~T. Neil, {\it {Review of strongly-coupled composite dark
  matter models and lattice simulations}},  {\em Int. J. Mod. Phys. A} {\bf 31}
  (2016), no.~22 1643004, [\href{http://arxiv.org/abs/1604.04627}{{\tt
  arXiv:1604.04627}}].

\bibitem{Bethe:1950jm}
H.~A. Bethe and C.~Longmire, {\it {The effective range of nuclear forces 2.
  photo-disintegration of the deuteron}},  {\em Phys. Rev.} {\bf 77} (1950)
  647--654.

\bibitem{Kaplan:1998tg}
D.~B. Kaplan, M.~J. Savage, and M.~B. Wise, {\it {A New expansion for
  nucleon-nucleon interactions}},  {\em Phys. Lett.} {\bf B424} (1998)
  390--396, [\href{http://arxiv.org/abs/nucl-th/9801034}{{\tt
  nucl-th/9801034}}].

\bibitem{Savage:1998ae}
M.~J. Savage, K.~A. Scaldeferri, and M.~B. Wise, {\it {N + p ---> d + gamma in
  effective field theory}},  {\em Nucl. Phys.} {\bf A652} (1999) 273--286,
  [\href{http://arxiv.org/abs/nucl-th/9811029}{{\tt nucl-th/9811029}}].

\bibitem{Krnjaic:2014xza}
G.~Krnjaic and K.~Sigurdson, {\it {Big Bang Darkleosynthesis}},  {\em Phys.
  Lett. B} {\bf 751} (2015) 464--468,
  [\href{http://arxiv.org/abs/1406.1171}{{\tt arXiv:1406.1171}}].

\bibitem{mccullough2}
W.~Detmold, M.~McCullough, and A.~Pochinsky, {\it {Dark Nuclei I: Cosmology and
  Indirect Detection}},  {\em Phys. Rev.} {\bf D90} (2014), no.~11 115013,
  [\href{http://arxiv.org/abs/1406.2276}{{\tt arXiv:1406.2276}}].

\bibitem{mccullough1}
W.~Detmold, M.~McCullough, and A.~Pochinsky, {\it {Dark nuclei. II. Nuclear
  spectroscopy in two-color QCD}},  {\em Phys. Rev.} {\bf D90} (2014), no.~11
  114506, [\href{http://arxiv.org/abs/1406.4116}{{\tt arXiv:1406.4116}}].

\bibitem{Hardy:2014mqa}
E.~Hardy, R.~Lasenby, J.~March-Russell, and S.~M. West, {\it {Big Bang
  Synthesis of Nuclear Dark Matter}},  {\em JHEP} {\bf 06} (2015) 011,
  [\href{http://arxiv.org/abs/1411.3739}{{\tt arXiv:1411.3739}}].

\bibitem{Hardy:2015boa}
E.~Hardy, R.~Lasenby, J.~March-Russell, and S.~M. West, {\it {Signatures of
  Large Composite Dark Matter States}},  {\em JHEP} {\bf 07} (2015) 133,
  [\href{http://arxiv.org/abs/1504.05419}{{\tt arXiv:1504.05419}}].

\bibitem{Mahbubani:2019pij}
R.~Mahbubani, M.~Redi, and A.~Tesi, {\it {Indirect detection of composite
  asymmetric dark matter}},  {\em Phys. Rev. D} {\bf 101} (2020), no.~10
  103037, [\href{http://arxiv.org/abs/1908.00538}{{\tt arXiv:1908.00538}}].

\bibitem{Pearce:2013ola}
L.~Pearce and A.~Kusenko, {\it {Indirect Detection of Self-Interacting
  Asymmetric Dark Matter}},  {\em Phys. Rev.} {\bf D87} (2013) 123531,
  [\href{http://arxiv.org/abs/1303.7294}{{\tt arXiv:1303.7294}}].

\bibitem{Foot:2014uba}
R.~Foot and S.~Vagnozzi, {\it {Dissipative hidden sector dark matter}},  {\em
  Phys. Rev. D} {\bf 91} (2015) 023512,
  [\href{http://arxiv.org/abs/1409.7174}{{\tt arXiv:1409.7174}}].

\bibitem{Pearce:2015zca}
L.~Pearce, K.~Petraki, and A.~Kusenko, {\it {Signals from dark atom formation
  in halos}},  {\em Phys. Rev.} {\bf D91} (2015) 083532,
  [\href{http://arxiv.org/abs/1502.01755}{{\tt arXiv:1502.01755}}].

\bibitem{Hardy:2014dea}
E.~Hardy, R.~Lasenby, and J.~Unwin, {\it {Annihilation Signals from Asymmetric
  Dark Matter}},  {\em JHEP} {\bf 07} (2014) 049,
  [\href{http://arxiv.org/abs/1402.4500}{{\tt arXiv:1402.4500}}].

\bibitem{Cirelli:2005uq}
M.~Cirelli, N.~Fornengo, and A.~Strumia, {\it {Minimal dark matter}},  {\em
  Nucl. Phys.} {\bf B753} (2006) 178--194,
  [\href{http://arxiv.org/abs/hep-ph/0512090}{{\tt hep-ph/0512090}}].

\bibitem{Blum:2016nrz}
K.~Blum, R.~Sato, and T.~R. Slatyer, {\it {Self-consistent Calculation of the
  Sommerfeld Enhancement}},  {\em JCAP} {\bf 1606} (2016), no.~06 021,
  [\href{http://arxiv.org/abs/1603.01383}{{\tt arXiv:1603.01383}}].

\bibitem{Hisano:2004ds}
J.~Hisano, S.~Matsumoto, M.~M. Nojiri, and O.~Saito, {\it {Non-perturbative
  effect on dark matter annihilation and gamma ray signature from galactic
  center}},  {\em Phys. Rev. D} {\bf 71} (2005) 063528,
  [\href{http://arxiv.org/abs/hep-ph/0412403}{{\tt hep-ph/0412403}}].

\bibitem{Hisano:2006nn}
J.~Hisano, S.~Matsumoto, M.~Nagai, O.~Saito, and M.~Senami, {\it
  {Non-perturbative effect on thermal relic abundance of dark matter}},  {\em
  Phys. Lett. B} {\bf 646} (2007) 34--38,
  [\href{http://arxiv.org/abs/hep-ph/0610249}{{\tt hep-ph/0610249}}].

\bibitem{Asadi:2016ybp}
P.~Asadi, M.~Baumgart, P.~J. Fitzpatrick, E.~Krupczak, and T.~R. Slatyer, {\it
  {Capture and Decay of Electroweak WIMPonium}},  {\em JCAP} {\bf 02} (2017)
  005, [\href{http://arxiv.org/abs/1610.07617}{{\tt arXiv:1610.07617}}].

\bibitem{Kaplan:2005es}
D.~B. Kaplan, {\it {Five lectures on effective field theory}},  10, 2005.
\newblock \href{http://arxiv.org/abs/nucl-th/0510023}{{\tt nucl-th/0510023}}.

\bibitem{Beane:2012vq}
{\bf NPLQCD} Collaboration, S.~R. Beane, E.~Chang, S.~D. Cohen, W.~Detmold,
  H.~W. Lin, T.~C. Luu, K.~Orginos, A.~Parreno, M.~J. Savage, and
  A.~Walker-Loud, {\it {Light Nuclei and Hypernuclei from Quantum
  Chromodynamics in the Limit of SU(3) Flavor Symmetry}},  {\em Phys. Rev.}
  {\bf D87} (2013), no.~3 034506, [\href{http://arxiv.org/abs/1206.5219}{{\tt
  arXiv:1206.5219}}].

\bibitem{Bethe:1949yr}
H.~A. Bethe, {\it {Theory of the Effective Range in Nuclear Scattering}},  {\em
  Phys. Rev.} {\bf 76} (1949) 38--50.

\bibitem{FERMI-LAT}
{\bf Fermi-LAT} Collaboration, M.~Ackermann et~al., {\it {Updated search for
  spectral lines from Galactic dark matter interactions with pass 8 data from
  the Fermi Large Area Telescope}},  {\em Phys. Rev.} {\bf D91} (2015), no.~12
  122002, [\href{http://arxiv.org/abs/1506.00013}{{\tt arXiv:1506.00013}}].

\bibitem{HESSDW}
{\bf HESS} Collaboration, H.~Abdalla et~al., {\it {Searches for gamma-ray lines
  and 'pure WIMP' spectra from Dark Matter annihilations in dwarf galaxies with
  H.E.S.S}},  {\em JCAP} {\bf 1811} (2018), no.~11 037,
  [\href{http://arxiv.org/abs/1810.00995}{{\tt arXiv:1810.00995}}].

\bibitem{Lefranc:2016fgn}
V.~Lefranc, E.~Moulin, P.~Panci, F.~Sala, and J.~Silk, {\it {Dark Matter in
  $\gamma$ lines: Galactic Center vs dwarf galaxies}},  {\em JCAP} {\bf 1609}
  (2016), no.~09 043, [\href{http://arxiv.org/abs/1608.00786}{{\tt
  arXiv:1608.00786}}].

\bibitem{Chun:2012yt}
E.~J. Chun, J.-C. Park, and S.~Scopel, {\it {Non-perturbative Effect and PAMELA
  Limit on Electro-Weak Dark Matter}},  {\em JCAP} {\bf 12} (2012) 022,
  [\href{http://arxiv.org/abs/1210.6104}{{\tt arXiv:1210.6104}}].

\bibitem{Fusco:2019zna}
P.~Fusco, {\it {OBSERVATION OF HIGH-ENERGY COSMIC PHOTONS WITH NEW-GENERATION
  SPACE TELESCOPES}},  {\em Frascati Phys. Ser.} {\bf 69} (2019) 138--143.

\bibitem{Ackermann:2015zua}
{\bf Fermi-LAT} Collaboration, M.~Ackermann et~al., {\it {Searching for Dark
  Matter Annihilation from Milky Way Dwarf Spheroidal Galaxies with Six Years
  of Fermi Large Area Telescope Data}},  {\em Phys. Rev. Lett.} {\bf 115}
  (2015), no.~23 231301, [\href{http://arxiv.org/abs/1503.02641}{{\tt
  arXiv:1503.02641}}].

\bibitem{Chen:2013bi}
J.~Chen and Y.-F. Zhou, {\it {The 130 GeV gamma-ray line and Sommerfeld
  enhancements}},  {\em JCAP} {\bf 1304} (2013) 017,
  [\href{http://arxiv.org/abs/1301.5778}{{\tt arXiv:1301.5778}}].

\bibitem{Bringmann:2016din}
T.~Bringmann, F.~Kahlhoefer, K.~Schmidt-Hoberg, and P.~Walia, {\it {Strong
  constraints on self-interacting dark matter with light mediators}},  {\em
  Phys. Rev. Lett.} {\bf 118} (2017), no.~14 141802,
  [\href{http://arxiv.org/abs/1612.00845}{{\tt arXiv:1612.00845}}].

\bibitem{Cirelli:2016rnw}
M.~Cirelli, P.~Panci, K.~Petraki, F.~Sala, and M.~Taoso, {\it {Dark Matter's
  secret liaisons: phenomenology of a dark U(1) sector with bound states}},
  {\em JCAP} {\bf 1705} (2017), no.~05 036,
  [\href{http://arxiv.org/abs/1612.07295}{{\tt arXiv:1612.07295}}].

\bibitem{Baldes:2017gzu}
I.~Baldes, M.~Cirelli, P.~Panci, K.~Petraki, F.~Sala, and M.~Taoso, {\it
  {Asymmetric dark matter: residual annihilations and self-interactions}},
  {\em SciPost Phys.} {\bf 4} (2018), no.~6 041,
  [\href{http://arxiv.org/abs/1712.07489}{{\tt arXiv:1712.07489}}].

\bibitem{HESSGC}
{\bf HESS} Collaboration, H.~Abdallah et~al., {\it {Search for $\gamma$-Ray
  Line Signals from Dark Matter Annihilations in the Inner Galactic Halo from
  10 Years of Observations with H.E.S.S.}},  {\em Phys. Rev. Lett.} {\bf 120}
  (2018), no.~20 201101, [\href{http://arxiv.org/abs/1805.05741}{{\tt
  arXiv:1805.05741}}].

\bibitem{Cirelli:2015bda}
M.~Cirelli, T.~Hambye, P.~Panci, F.~Sala, and M.~Taoso, {\it {Gamma ray tests
  of Minimal Dark Matter}},  {\em JCAP} {\bf 1510} (2015), no.~10 026,
  [\href{http://arxiv.org/abs/1507.05519}{{\tt arXiv:1507.05519}}].

\bibitem{Schiff}
L.~I. Schiff, {\em Quantum Mechanics (3rd edition)}.
\newblock McGraw-Hill, 1968.

\bibitem{Ershov:2011zz}
S.~Ershov, J.~Vaagen, and M.~Zhukov, {\it {Modified variable phase method for
  the solution of coupled radial Schrodinger equations}},  {\em Phys. Rev. C}
  {\bf 84} (2011) 064308.

\bibitem{Beneke:2014gja}
M.~Beneke, C.~Hellmann, and P.~Ruiz-Femenia, {\it {Non-relativistic pair
  annihilation of nearly mass degenerate neutralinos and charginos III.
  Computation of the Sommerfeld enhancements}},  {\em JHEP} {\bf 05} (2015)
  115, [\href{http://arxiv.org/abs/1411.6924}{{\tt arXiv:1411.6924}}].

\end{thebibliography}\endgroup

\end{document}